# Frontlines and faultlines: How the Russo-Ukrainian conflict reshapes the landscape of scientific research

Yang Ding

Business School, The University of Edinburgh, Edinburgh EH8 9JS, United Kingdom

Corresponding Author: Yang Ding (yang.ding@ed.ac.uk)

**Abstract**

Geopolitical conflict poses significant challenges to research and innovation policy by disrupting scientific systems and talent mobility. This study analyzes the impact of the Russia-Ukraine conflict, particularly the escalations in 2014 and 2022, on the academic landscapes of both countries. We collected data on academic publications from 2000 to 2023, including 272,924 publications linked to Ukrainian scholars across 425 universities and facilities and 210,765 scholars in Ukraine, and 1,533,758 publications linked to 1,968 universities and facilities and 887,624 scholars in Russia. We also gathered collaboration data between Ukraine, Russia, and 193 other countries and regions. We tracked scholar migration and domestic research activity, compared research topics between stayed active scholars and departed scholars, and examined the evolution of international collaboration networks. We find significant migration following both the 2014 and 2022 events, with a particularly severe impact on Ukraine in terms of both talent loss and a sharp decline in domestic research visibility. Migrated Ukrainian scholars increased activity in internationalized basic science fields (e.g., particle physics), while stayed active scholars focused more on applied and strategic areas related to national challenges, reconstruction, and resilience amid conflict and sanctions. Both departed and stayed active scholars experienced decreased output in resource-dependent fields, particularly medical research. International collaboration networks were reshaped, with traditional ties (Russia-West, Ukraine-Russia) dissolving and new collaborations (Russia–neighboring countries, Ukraine–West) rising. Ukrainian scholars migrating to other countries face challenges assuming key research roles, though some smaller host countries' academic communities showed a trend toward leadership positions. Russian scholars saw a decline in research prominence across most countries, affected by international sanctions. These findings reveal how conflict disrupts national scientific capacity, fractures global research networks, and affects individual academic careers, highlighting the need for targeted policies to support vulnerable academic communities and preserve international scientific cooperation during crises.

# 1 Introduction

War profoundly and enduringly impacts societies, economies, and the very foundation of human scientific and knowledge production. Modern scientific activity thrives on the concentration of human capital, collaborative networks spanning borders and institutions, and a stable environment supported by consistent resource investment (Kitajima & Okamura, 2025; Ponomariov & Boardman, 2010). However, this inherent reliance renders the scientific system particularly vulnerable to external shocks, with geopolitical conflict standing out as one of the most destructive forces. Such conflict not only directly endangers researchers and devastates crucial infrastructure but also systematically undermines the structure and function of scientific social networks through altered political landscapes, imposed sanctions, and disrupted funding streams (McLean & Whang, 2020; Wolfe, 2013; Zhang et al., 2024). As the critical carriers of knowledge and skills, the mobility of human capital, while vital, paradoxically exacerbates the system's fragility in times of war, leading to talent drain and a further weakening of the scientific base (Docquier & Rapoport, 2012; Gibson & McKenzie, 2011). Consequently, geopolitical conflict not only distorts the priorities of knowledge creation and fundamentally disrupts the normal trajectory of academic talent but also severely shocks existing scientific institutions and norms, compelling the academic community to undergo arduous adaptation within highly uncertain and resource-constrained environments.

The ongoing Russia-Ukraine conflict, which has escalated progressively since 2014 and erupted into full-scale war in 2022, constitutes an unprecedented systemic disruption to a significant scientific region. The protracted nature and complexity of this conflict make it a critical case study for in-depth exploration of how large-scale geopolitical shocks operate on the core elements of national innovation systems—particularly their human capital base and knowledge collaboration networks—triggering profound structural changes. Recent data suggests that by 2022, Ukraine’s gross domestic expenditure on R&D declined by 38.5%, and approximately 12% of its scholars have been displaced (Hladchenko, 2025, 2026). Understanding the mechanisms and consequences of this shock is essential for building more resilient global science systems and formulating human capital and research policies during times of crisis, and is equally crucial for receiving countries in understanding and addressing the challenges of integrating displaced scholars and their talent and knowledge, and in seeking their effective integration (Sweetman, 2025).

While existing work offers preliminary understanding of the potential impacts of war and conflict on academia, utilizing survey methods to explore dimensions such as gender (Shkoda & Ovchynnikova, 2025), mobility (de Rassenfosse et al., 2023; Jaroszewicz et al., 2025), institutions (Plastun & Kozmenko, 2025; Sweetman, 2025; Zayachuk, 2024), and productivity (Zhang et al., 2024), comprehensive, in-depth, and quantitative evidence derived from large-scale behavioral data is still needed to fully understand how this geopolitical shock specifically translates into structural transformation within the academic system (Gaind et al., 2022). Bibliometric tracking has recently revealed a significant shift in destination countries for Ukrainian scholars, with Russia being overtaken by Germany, Poland, and China as primary hubs (Hladchenko, 2025). Existing studies provide limited systematic insight into key aspects: How does conflict quantitatively alter the scale and dynamics of scholar migration as well as the continued activity of those who remain? How does this forced or voluntary reshaping of human capital lead to divergence and evolution in the priority areas of knowledge production? How are the core networks of international scientific collaboration fundamentally reshaped? And, how does the academic role and integration of displaced scholars change within new institutional and network environments in host countries? Answering these questions requires moving beyond perceptions or localized snapshots to analyze long-term scholar trajectories and collective interaction patterns.

This study aims to provide a systematic quantitative analysis of the dynamic changes within the academic systems of Ukraine and Russia between 2000 and 2023 in the context of the ongoing conflict. We leverage OpenAlex by tracking scholars through their works and institutional affiliations over time. We offer quantitative evidence of conflict-driven scholar migration patterns and assess the impact of the 2014 and 2022 key events on migration dynamics. We further analyze the impact of the conflict on the direction of knowledge production within the Russian and Ukrainian academic communities, differentiating the diverging trajectories of research topics between scholars who remained and those who migrated. Simultaneously, we explore how the conflict reshaped the international academic collaboration networks of both countries. Finally, we examine the integration of Departed Scholars into the academic systems of major host countries, analyzing changes in their academic roles in co-authored publications to reveal shifts in their status. The findings of this study aim to provide an important empirical basis for understanding the mechanisms through which geopolitical conflict disrupts the core elements of national innovation systems simultaneously driving talent exodus and suppressing domestic scientific vitality, offering insights for public policies focused on human capital retention, supporting displaced talent, and maintaining international scientific collaboration in crisis environments.

## 2 Related Literature

One of the most direct and visible impacts of extreme events like war on the academic community is the phenomenon of "brain drain," or the loss of skilled talent (Science News Staff, 2024). Historical studies have extensively documented large-scale migrations of scientists during periods of conflict or political turmoil and analyzed their potential positive impacts on the innovation capacity of receiving countries (Borjas & Doran, 2012; Gaind et al., 2022; Moser et al., 2014) However, for the talent-exporting countries, the outflow of highly skilled individuals represents a significant loss, directly weakening domestic research capacity and potentially generating multiplier effects by reducing local collaborators and hindering the training of the next generation of researchers (Karmadonova, 2023; Popović, 2025; Schiermeier, 2014).

Concerning the Russia-Ukraine conflict, survey-based research has focused on its impact on the Ukrainian scientific community, finding that a considerable proportion of Ukrainian scientists were forced to leave the country after the 2022 invasion, including many of the most active researchers(de Rassenfosse et al., 2023). The migrated scholars often face challenges in their new environments, such as precarious contract situations, and their degree of integration into new institutions and ability to secure independent research opportunities can depend heavily on the receiving context (Challoner & Forget, 2011). Scholars who remain in conflict zones also face severe challenges, including drastic reductions in available research time and the destruction of or inability to access critical research resources and infrastructure (Turkmen et al., 2024). Furthermore, the physical and psychological burdens of war impose immense challenges on remaining scholars (Dobiesz et al., 2022; Turkmen et al., 2024). Long-term exposure to conflict environments significantly negatively impacts the mental and physical health of academic staff, including university faculty and students, increasing fear, burnout, and psychological stress responses (Kurapov et al., 2024). In tandem with the displacement in Ukraine, the Russian scientific community has also witnessed a significant exodus; recent bibliometric evidence indicates a substantial decline in the net migration rate of Russian-affiliated researchers, with the country losing approximately 0.8% of its active scholarly population annually across nearly all research fields (Lovakov, 2025).

This focus on displacement is part of a broader re-examination of regional academic mobility, which was previously characterized as a maturing model of "brain circulation" rather than simple depletion. Indeed, the foundational work of Subbotin and Aref (2020) demonstrated that, prior to recent escalations, the Russian scientific system had achieved a relatively balanced international exchange of talent across diverse disciplines. Building on this established circulation framework, our study extends the analysis by investigating how extreme geopolitical shocks specifically alter the academic roles and thematic trajectories of these mobile scholars.

Beyond affecting the individual circumstances of scholars, conflict profoundly alters the substantive content and organizational modes of research activities (Guariglia et al., 2025). In the context of conflict and sanctions, national research priorities may shift, with resource allocation tilting towards specific strategic or applied fields, while areas dependent on international collaboration and high-end equipment may face decline (Dobrovidova, 2023). International academic collaboration networks are vital conduits for knowledge diffusion and innovation, but geopolitical tensions directly threaten their stability. For instance, existing observations and preliminary analyses suggest a decline in international research collaboration involving Russia after the outbreak of the Russia-Ukraine conflict (Dyachenko et al., 2024; Plackett, 2022; Poskett & Shaw, 2022; Science News Staff, 2025). Russia dropped from being the top collaborating partner for National Academy of Sciences of Ukraine (NASU) to ninth place since 2022 (Hladchenko, 2026). This is more than a simple interruption of collaboration; it can lead to fragmentation and long-term disconnection in relevant knowledge domains. However, scholar migration is not entirely negative; migrated scientists can sometimes act as "bridges" connecting their home and host countries, fostering the formation of international research networks and providing opportunities for home-country colleagues to access new knowledge and resources (Jonkers & Cruz-Castro, 2013). This "bridge" effect or the subsequent return of scholars can, to some extent, buffer the negative impact of brain drain (Özbilgin et al., 2025).

While existing research offers some preliminary understanding of the impact of conflict on collaboration and overall research output through surveys and macro-level output analysis (Lem, 2022; Mäkinen, 2023; Noorden, 2023; Schiermeier, 2020), detailed longitudinal analysis is largely absent regarding the specific structural transformations within academic systems. Although restrictions and limitations on Russian scholars' international collaboration due to sanctions and the war have been widely discussed, the contrasting situation for Ukrainian scholars has received significantly less research attention (Gaind et al., 2022; Turkmen et al., 2024). Numerous countries and organizations have offered policy and research funding support for hosting displaced Ukrainian scholars (Plackett, 2025; Stone, 2024). This support, while vital, might inadvertently create a blind spot, causing us to overlook the complexities of their actual research collaboration and integration status in host countries. Therefore, few studies specifically examine the resulting changes in collaboration opportunities and academic roles for Ukrainian scholars, both for those who remained in the country and those who migrated abroad. Analysis of how the actual status and contribution of migrated scholars change after integrating into host country academic systems (particularly as reflected in their roles within research teams) is crucial for both the individual career development of these scholars and the potential for their future return to their home country to rebuild its science (Duszyński et al., 2022; Reekie, 2023; Shulga, 2023). Furthermore, detailed evidence is lacking on how the research topics of both Russian and Ukrainian scholars specifically evolved before and after the conflict. The reshaping of their primary international collaboration partners also remains underexplored. In-depth exploration of these aspects offers a more nuanced understanding of how geopolitical conflict, through the critical channel of talent flow, reshapes a nation's scientific innovation capacity and its position within the global academic network.

## 3 Data and Methods

### 3.1 Data Source and Collection

Our research data are derived from OpenAlex, a comprehensive, open-access index of global scholarly research encompassing entities such as publications, authors, institutions, and concepts, along with their interconnections (Céspedes et al., 2025). OpenAlex serves as a significant successor to Microsoft Academic Graph and is a powerful tool for analyzing macro-level science activity patterns, collaboration networks, and institutional affiliations.

We collected data on scientific publications from the OpenAlex platform spanning the period from 2000 to 2023. We further extracted information on all authors, their affiliated institutions, and related metadata associated with these publications. Following initial collection and deduplication, our dataset includes 272,924 scientific publications associated with Ukrainian scholars, involving 425 universities and research institutions within Ukraine and 210,765 individual scholars. For Russia, the dataset comprises 1,533,758 scientific publications, involving 1,968 universities and research institutions within Russia and 887,624 scholars. Each scholar was assigned a unique author identifier. This mechanism ensures that the identifier remains consistent regardless of how a scholar changes affiliations or migrates to different countries throughout their career, enabling accurate tracking of individuals' academic activities across time and institutions. Scholars who published and conducted research continuously within Ukraine or Russia for at least a decade prior to 2014 were identified as representative scholars for Ukraine or Russia, respectively.

Furthermore, these publications reflect extensive international collaboration, involving co-authors from 193 countries and regions worldwide. These publications co-authored with Ukrainian or Russian scholars are associated with 23,274 universities and research institutions located outside Ukraine and Russia, and involve 1,091,892 collaborating scholars. While OpenAlex provides superior coverage of regional and non-English publications, we acknowledge inherent challenges in metadata completeness and institutional linking common to automated aggregators (Zhang et al., 2024). A detailed technical disclosure of global metadata missingness and our internal quality audits for the target populations is provided in **Supplementary Note 1 and Note 2**.

### 3.2 Data Structure and Processing

The data utilized in this study are structured and integrated, recording the relationships between authors, publications, and their affiliated institutions. Key metadata extracted from OpenAlex include a unique Work ID for each scientific publication; the author's position within the publication's author list, which is used to identify roles such as first author, middle author, and last author; a unique Author ID for each scholar, essential for tracking individual publication activity and collaborations over time; a unique Institution ID for each university or research institution; display names for scholars and institutions; the country code and geographical coordinates (latitude and longitude) of institutions, used to define national affiliation and identify international collaboration patterns; the institution type (e.g., university, research institution, company, etc.); the publication year of the scientific work, which is crucial for conducting time-series and trend analyses; and research topics associated with each scientific publication, obtained from OpenAlex's internal classification system.

To define national affiliation and identify international collaboration patterns, we utilized the country code and geographical coordinates (latitude and longitude) of institutions. To address potential biases arising from multiple affiliations, joint appointments, or temporary visiting positions, we implemented a "Domestic Anchor Protocol." This protocol dictates that a migration event is only recorded upon the complete severance of domestic institutional links, thereby providing a conservative lower-bound estimate of talent mobility (see **Supplementary Note 3**). A quantitative audit (see **Supplementary Note 4**) confirmed a low profile fragmentation rate (upper bound 5.27%), ensuring the statistical robustness of our longitudinal analysis and macro-level findings.

### 3.3 Time Series Analysis

Based on scholars' publication affiliation information across the study period (2000-2023), we classified the Ukrainian and Russian scholar populations into two broad categories: "Stayed Active Scholars" and "Departed Scholars." The foundational sample for this classification analysis was restricted to scholars who had at least one publication affiliated with an institution in their home country (Ukraine or Russia) prior to 2014, and who also had at least one scientific publication in 2014 or later. This screening criterion ensures that we could observe the publishing activities and affiliation changes of scholars before and after the key conflict events and the full-scale war, allowing us to compare the trajectories of scholars who remained versus those who left.

A scholar $a$ was classified as a stayed active scholar if the country codes of the affiliated institutions for all their publications in 2014 and later remained exclusively within their home country $H_a$. This classification is determined ex post based on each individual's complete publication record from 2014 to 2023, ensuring each scholar is assigned to one and only one category throughout the observation window. Then,

$$a \in S_{stayed} \Leftrightarrow a \in S_{Classified} \wedge AffCountrySet\left(a,(2014,2023]\right) \subseteq \left\{H_a\right\} \tag{1}$$

Conversely, a scholar $a$ was classified as a departed scholar if the set of country codes for the affiliated institutions on all their publications in 2014 and later included only country codes outside their home country $H_a$. Specifically, this group includes scholars who recorded an irreversible relocation without returning during the study period. Then,

$$a \in S_{departed} \Leftrightarrow a \in S_{Classified} \wedge H_a \notin AffCountrySet\left(a, (2014, 2023]\right) \tag{2}$$

This binary classification allows us to conduct comparative analyses of the academic trajectories and publication patterns of scholars who remained within the domestic institutional context versus those who relocated to international environments.

To quantify and assess the impact of the two key events—the 2014 conflict commencement and the 2022 full-scale war escalation—on the patterns of scholar outflow from Ukraine and Russia, we employed Interrupted Time Series (ITS) analysis (McDowall et al., 1980). ITS is a robust quasi-experimental design particularly suitable for evaluating the effect of specific interventions or events on the level and trend of an outcome variable over time, while accounting for pre-existing temporal trends (Turner et al., 2019). We employ heterogeneous temporal indexing for the outcome variable $Y_t$ to accurately capture the distinct dynamics of the two cohorts. For Departed Scholars, $Y_t$ represents event based flows, recording each migrated scholar only once in the year of their first permanent departure. For Stayed Active Scholars, $Y_t$ represents activity based presence, counting scholars only in the specific years they have at least one indexed publication. This approach accounts for the inherent asynchrony and intermittency of research cycles, where a domestic scholar may remain in the system but only appear in the database during their active publishing years. We constructed separate ITS models for Ukrainian and Russian scholars. The models included two main intervention points corresponding to the major conflict escalations: 2014 (the first shock) and 2022 (the second shock).

The basic form of the ITS model employed is:

$$Y_t = \beta_0 + \beta_1 T_t + \beta_2 L_{1t} + \beta_3 Post^{2014} T_t + \beta_4 L_{2t} + \beta_5 Post^{2022} T_t + \varepsilon_t \quad (3)$$

where $Y_t$ represents the number of scholar outflows in year $t$. $Time_t$ is a continuous time counter starting from 1 in 2000 ($t = 1$ for 2000, $t = 2$ for 2001, and so on). $L_{1t}$ is a binary indicator variable for the first intervention point, taking a value of 1 for $t \geq 2014$ and 0 otherwise. $\beta_1$ captures the baseline annual trend in scholar outflow before the 2014 conflict. $\beta_2$ represents the immediate change in the level of scholar outflow occurring in 2014 due to the first conflict shock. $Post^{2014} T_t$ is a time variable measuring the number of years elapsed since the first intervention, defined as $t - 2013$ for $t \geq 2014$ and 0 otherwise. $\beta_3$ estimates the change in the annual trend of scholar outflow *after* the 2014 conflict compared to the pre-2014 trend. $L_{2t}$ is a binary indicator variable for the second intervention point, taking a value of 1 for $t \geq 2022$ and 0 otherwise. $\beta_4$ represents the immediate change in the level of scholar outflow occurring in 2022 due to the outbreak of the full-scale war. $Post^{2022} T_t$ is a time variable measuring years elapsed since the second intervention, defined as $t - 2021$ for $t \geq 2022$ and 0 otherwise. $\beta_5$ estimates the change in the annual trend of scholar outflow *after* the 2022 conflict compared to the trend before 2022. $\varepsilon_t$ is the error term. By analyzing the coefficients and statistical significance of $\beta_2$, $\beta_3$, $\beta_4$, and $\beta_5$, we can quantify the short-term (level shifts) and long-term (trend changes) impacts of the two conflict events on scholar outflow from Russia and Ukraine.

Our approach primarily relies on analyzing the evolution of affiliated institution information listed in scientific publications over time to track scholars' geographical mobility and changes in institutional affiliation. This method is effective in revealing patterns of scholarly activity across geographical borders. However, it is important to acknowledge that the typical lifecycle of a scientific publication—from research conception, writing, submission, peer review, to final acceptance and online publication—introduces an inherent time lag. It commonly takes several months, or even a year or longer, from the completion of a research project until a publication reflecting updated affiliation information is ultimately indexed in academic databases. This implies a noticeable delay between the actual time a scholar physically migrates or changes institutional affiliation and the point in time when this new affiliation appears in their publications. For example, a scholar might have effectively left Ukraine or Russia and secured a research position in a new country in 2014, but a publication listing their new institution as the affiliation is likely to be published and enter databases like OpenAlex only in 2015 or later. To assess the robustness of our main ITS analysis results to varying degrees of this potential publication lag, we conducted supplementary sensitivity analyses. Specifically, we re-estimated the ITS models by shifting the key intervention points forward (lagging) by one year, to simulate the effects of different publication lag durations. For the 2014 intervention, we tested scenarios where the intervention point was set to 2015. For the 2022 intervention, given the dataset's end year of 2023, we primarily tested the scenario where the intervention point was set to 2023 (simulating a one-year lag). The results of these robustness checks, including the estimated coefficients and their statistical significance under different lagged intervention point assumptions, are presented in **Supplementary Note 5**.

In addition to analyzing scholar migration, we conducted several extended analyses to investigate the impact of geopolitical conflict on research topics, international collaboration, and academic roles. Leveraging the topic information associated with each publication, we tracked the changes in the relative activity levels (measured by publication output within specific topics) of scholars who permanently departed their home country (not having returned by the end of the study period) versus those who remained. We estimated the trend in their engagement with representative research topics by analyzing the slope of topic-specific publication counts over time using linear regression. Second, we assessed the evolution of the international academic collaboration landscape for both country groups. This involved examining the annual number of distinct collaborating countries and analyzing changes in the proportion of collaborations with their most frequent international partners to identify potential network expansion or restructuring. Using the author position information available in the publication metadata, we analyzed the proportion of publications where migrated scholars in major receiving countries were listed as first authors, middle authors, or last authors over time. This analysis serves to shed light on their evolving academic status and capacity to lead research projects within their new institutional environments. Crucially, our framework is methodologically robust to potential undercounting arising from metadata incompleteness. Since the model estimates relative shifts in intercepts and trajectories rather than relying on absolute volume, the identified structural breaks remain valid as long as the rate of metadata missingness does not fluctuate in synchronization with the intervention points.

To quantify international collaboration, we adopted a binary whole-counting approach at the national level. A publication is recorded as a single link between the home country and a partner nation if at least one institutional affiliation from that partner is present. Multiple co-authors from the same country do not increase the weight for that specific publication. This method ensures the network reflects the breadth of international reach and avoids skewing results by the team sizes of large-scale projects. To ensure the statistical stability of the collaboration metrics, we excluded hyper-authored publications with more than 100 contributors (see **Supplementary Note 6**).

To further evaluate the robustness of our findings regarding the impact of conflict on research topic changes, we conducted a placebo test using the Doubly Robust Difference-in-Differences (DRDID) method (Sant'Anna & Zhao, 2020). This analysis, detailed in **Supplementary Note 7**, involves constructing a placebo dataset where scholars within each country subgroup were randomly assigned to treatment and control groups. These placebo groups simulate hypothetical scenarios where Ukraine and Russia were not exposed to conflict, allowing us to assess for spurious effects by comparing them against the actual treatment groups.

# 4 Results

## 4.1 Conflict-Driven Scholar Migration

Figure 1a illustrates the distinct migration patterns of Ukrainian and Russian scholars following the outbreak of the conflict in Crimea in 2014 and the escalation to full-scale war in 2022. Scholars departing from Ukraine primarily moved to research institutions in neighboring European countries. In contrast, the outflow of Russian scholars was more geographically dispersed, with destinations spanning multiple continents, including North America, Europe, and Asia. Figure 1b details the primary destination countries for scholars from both nations. Notably, a significant portion of scholars departing Ukraine chose Russia as a destination, likely due to close geographical, cultural, and linguistic ties. Another important factor is the considerable number of scholars of Russian nationality returning to their home country from Ukraine. Scholars leaving Russia, on the other hand, primarily migrated to countries such as the United States, China, Germany, and the United Kingdom. The initial cross-border migration of Ukrainian and Russian scholars in 2023, displaying trends consistent with the post-2014 period, is depicted in figures within **Supplementary Note 8**.

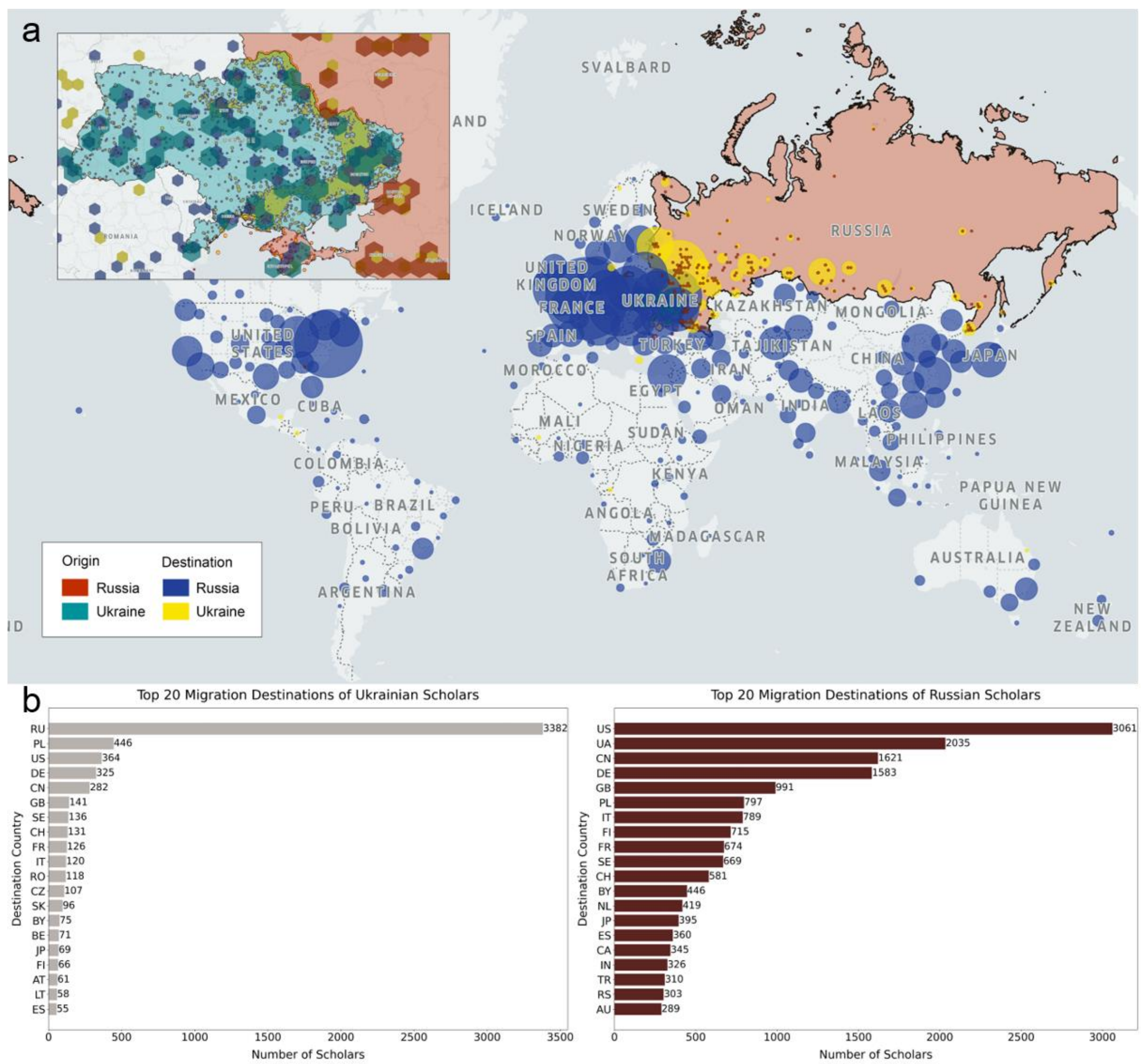

**Figure 1. Scholar migration characteristics and destination distribution for Russia and Ukraine.** Figure 1a shows the schematic of Russian and Ukrainian scholar migration patterns. Figure 1b shows the distribution of primary destination countries for departed Russian and Ukrainian scholars.

Figure 2a displays the changes in the annual count of Stayed Active Scholars as well as the frequency of permanent migration events from Russia and Ukraine. It reveals a sharp increase in the number of migration events after 2014. Concurrently, the number of Stayed Active Scholars who remained active within their home countries began a rapid decline after 2021. This decrease reflects a significant reduction in the annual scientific visibility and productivity of the domestic workforce, indicating that conflict not only drives physical departure but also suppresses the research activity of those who remain. In 2023 alone, over 1,600 Ukrainian scholars permanently left their home country and published research in other countries, while the number of scholars who departed Russia and published abroad in the same year reached over 5,500. Notably, a visible increase in migration events is observed starting around 2020, prior to the 2022 escalation. This rise aligns with the statistically significant trend acceleration ($\beta_3$) identified in our analysis, reflecting the cumulative as well as intensifying pressures on the scientific communities following the 2014 conflict. This pre war surge likely represents the anticipatory migration of forward looking scholars responding to the deteriorating geopolitical climate. Furthermore, from 2021 to 2023, the number of scholars who remained active in academic research within Ukraine decreased by 13,000, and this figure exceeded 60,000 for Russia.

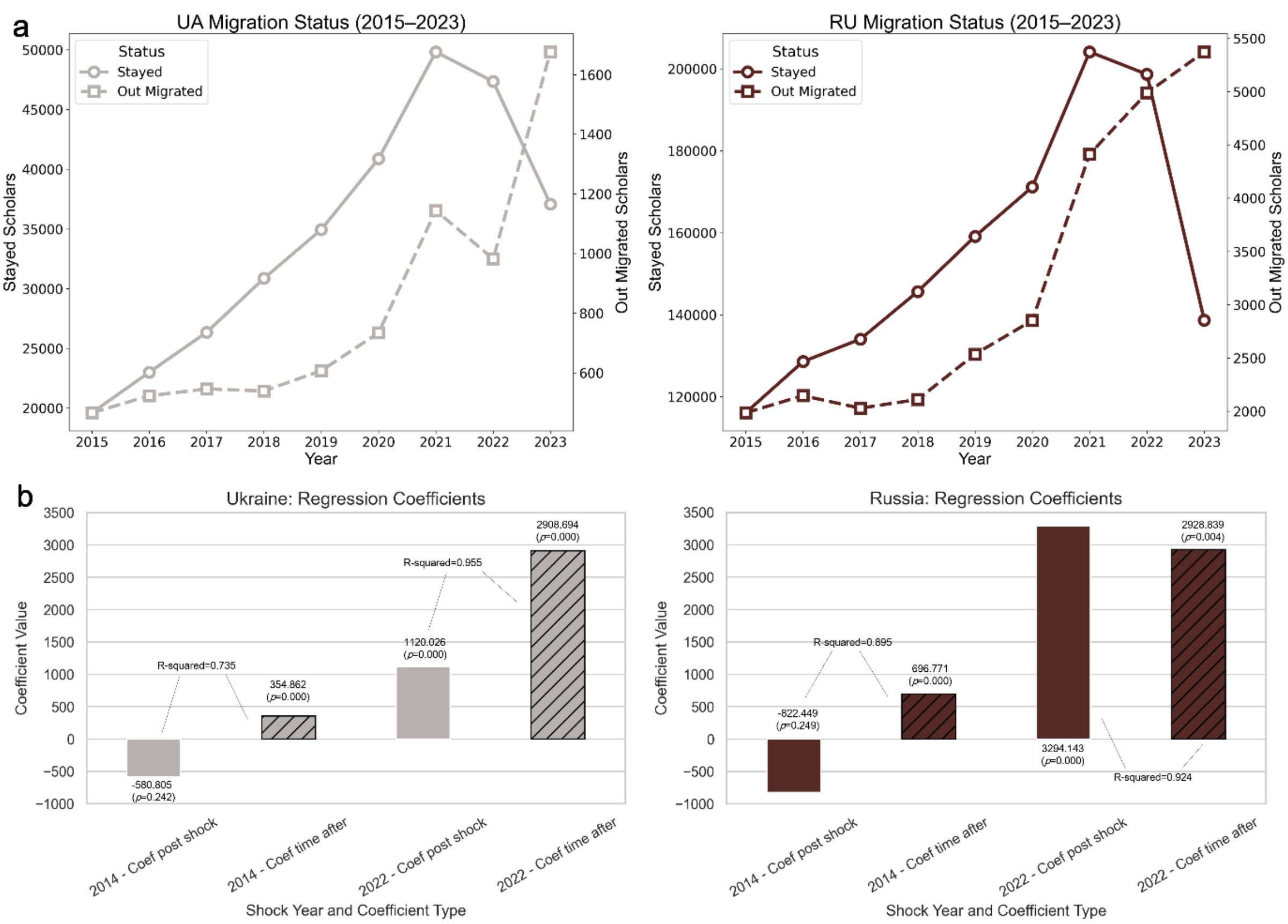

**Figure 2**. Quantitative impact analysis of conflict on Russian and Ukrainian scholar outflow. Figure 2a represents annual number of departed scholars and domestic remaining scholars in Russia and Ukraine.

Figure 2b shows the ITS time series analysis results for the impact of conflict on outflow. The robustness results are highly consistent with the main model's findings (intervention points set at 2014 and 2022), indicating that the study's main conclusions regarding the conflict's impact on scholar outflow are robust to potential publication lag effects.

Figure 2b presents the results of the ITS analysis, further quantifying and revealing the differential impacts of the 2014 and 2022 shock events on scholar migration incidence from Ukraine and Russia. For Ukraine, the 2014 shock (conflict onset) did not lead to a statistically significant immediate change in the level of scholar outflow ($coef. = -580.805$, $p = 0.242$). However, following 2014, the trend of Ukrainian scholar outflow showed a statistically highly significant acceleration ($coef. = 354.862$, $p = 0.000$). This indicates that although there was no immediate mass exodus, the protracted impact of the conflict spurred a continuous increase in the rate of initial migration events in the subsequent years. Compared to 2014, the 2022 full-scale war had a far more drastic and immediate impact on Ukrainian scholar outflow. The analysis results show a statistically highly significant immediate increase in the level of Ukrainian scholar outflow after the 2022 shock ($coef. = 1120.026$, $p = 0.000$). More importantly, the trend of outflow after 2022 also exhibited a statistically highly significant and substantial acceleration ($coef. = 2908.694$, $p = 0.000$). This clearly demonstrates that the 2022 events triggered a massive, immediate outflow of Ukrainian scholars, which maintained a rapid growth trajectory in the following period, with an impact far exceeding that of 2014.

For Russia, the scholar outflow pattern is similar to Ukraine's in some respects but shows key differences. Scholars may also have faced risks associated with the war and political instability, accelerating their departure. Despite limited direct combat within Russia, the potential economic and political risks resulting from international sanctions imposed on Russia also expedited this brain drain process. The 2014 shock similarly did not result in a statistically significant immediate change in the level of Russian scholar outflow ($coef. = -822.449$, $p = 0.249$), but the subsequent trend of scholar outflow also showed a statistically highly significant acceleration ($coef. = 696.771$, $p = 0.000$). This suggests that the 2014 events had a long-term trend-setting impact on Russian scholar outflow. The 2022 full-scale war had a dramatically significant impact on Russian scholar outflow. Following the 2022 shock, the level of Russian scholar outflow experienced a statistically highly significant surge ($coef. = 3294.143$, $p = 0.000$). Notably, the magnitude of this immediate increase was even greater than that observed for Ukraine in 2022. Furthermore, the trend of Russian scholar outflow after 2022 maintained a statistically significant acceleration ($coef. = 2928.839$, $p = 0.004$). This indicates that the 2022 events led to a massive, immediate outflow of Russian scholars, which continued to accelerate subsequently, signifying a very profound impact.

The ITS analysis results quantitatively reveal the immense shock of the two conflict events on the active scientific talent pool in both Russia and Ukraine, highlighting the 2022 full-scale war as a critical turning point that triggered an unprecedented wave of scholar exodus. Russia's total number of universities, research institutions, and scholars before the war significantly exceeded Ukraine's. The long-term outflow trend coefficient after the 2022 full-scale war is remarkably similar for Ukraine and Russia (2908.694 vs. 2928.839). On one hand, this implies that a talent outflow of this magnitude is an almost devastating blow to Ukraine's smaller research environment and scholar population. On the other hand, Russia has strengthened cooperation with neighboring countries in recent years and established numerous scholarship and research funding programs to attract students and scholars from surrounding countries for education or scientific research (Istyagina-Eliseeva et al., 2022; Schiermeier, 2020). International sanctions have also led to the return of some Russian scholars who had previously moved to other countries (Plackett, 2024). The influx and return of these scientific talents may partially compensate for the gap created by the outflow, a factor that Ukraine, facing direct conflict, currently finds difficult to match.

### 4.2 Research Topic Shifts and Divergence

Figure 3 reveals a significant divergence in research fields and activity among Ukrainian scholars since 2014, particularly influenced by the escalating Russia-Ukraine conflict. For the group of Ukrainian scholars who left their home country, there is a trend towards polarization in research directions. At one end, demonstrating increased research vitality (positive slopes), certain frontier basic science fields show robust growth, such as "Search for Quark-Gluon Plasma in Heavy-Ion Collisions," "Particle Physics and High-Energy Collider Experiments," and "Advancements in Particle Detector Technology." This trend may benefit from the highly internationalized research networks and large collaborative projects (like the European Organization for Nuclear Research) characteristic of these fields (CERN, 2022). Scholars who left Ukraine could more easily integrate into and utilize these global resources and infrastructures, thereby maintaining or even accelerating their research output. Additionally, "Assessment of Sustainable Development Indicators and Strategies" and some economics-related topics also show positive slopes, potentially reflecting scholars' continued focus on global issues or macro-level problems closely linked to Ukraine's future development in their new environments (Berglöf & Rashkovan, 2023).

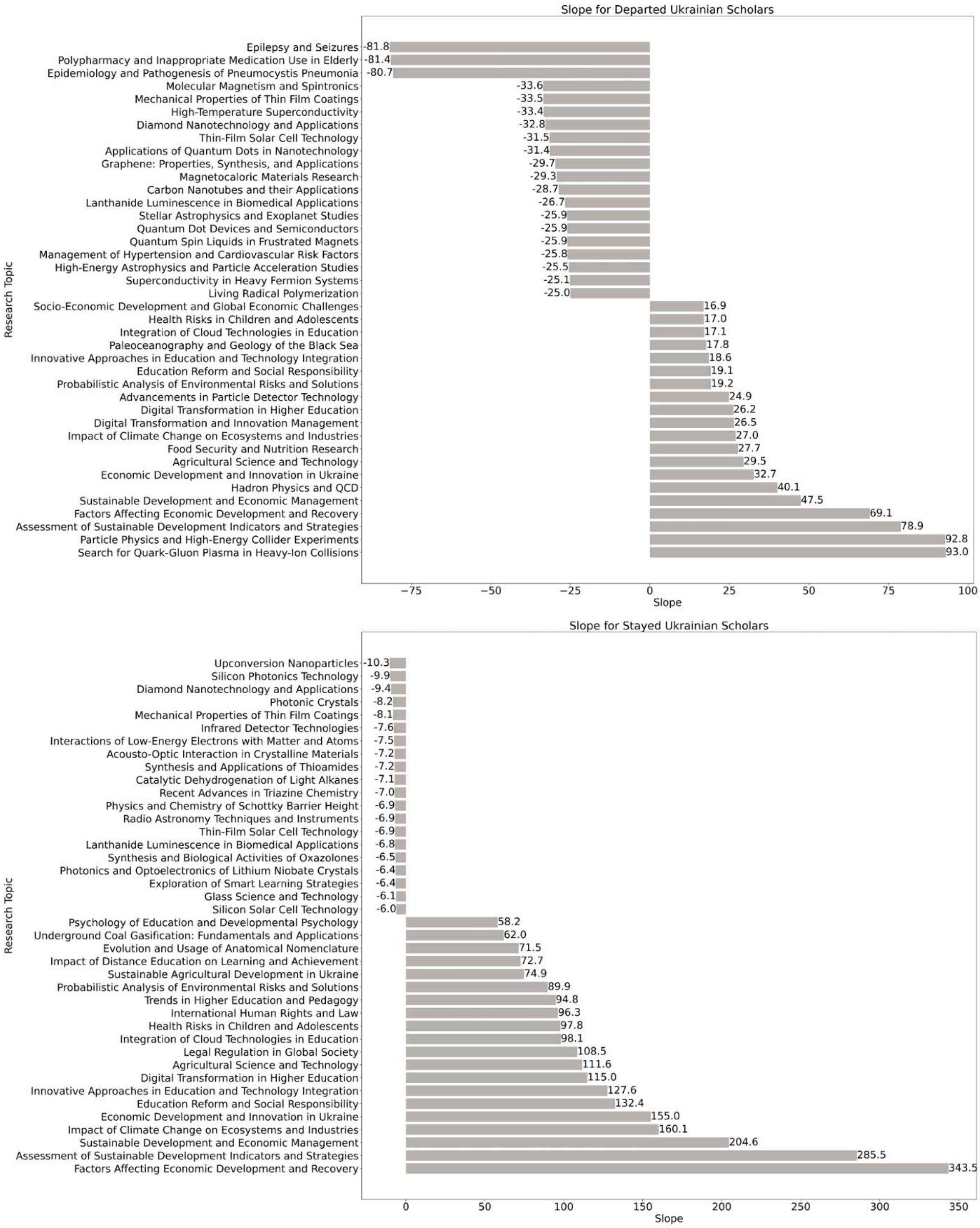

**Figure 3. Trends in research topics over time for Ukrainian Scholars (Departed vs. Stayed).**

However, at the other end, showing declining research vitality (negative slopes), research activities in certain fields experienced alarming sharp decreases, particularly in some medical-related areas such as "Epilepsy and Seizures," "Polypharmacy and Inappropriate Medication Use in Elderly," and "Epidemiology and Pathogenesis of Pneumocystis Pneumonia." Some condensed matter physics and materials science topics, such as "Molecular Magnetism and Spintronics," "Mechanical Properties of Thin Film Coatings," and "High-Temperature Superconductivity," also show clear negative slopes. This downward trend strongly suggests that research in these areas is highly dependent on specific experimental equipment, laboratory environments, or clinical resources. The physical migration of scholars led to their inability to access these critical resources, severely disrupting research activities. The relatively high absolute values of the negative slopes indicate that these fields suffered a significant impact among the group of departed scholars.

In contrast, the group of Ukrainian scholars who remained in their home country demonstrated greater resilience and growth in more applied topics closely related to Ukraine's domestic socio-economic development. Topics with high positive slopes include "Factors Affecting Economic Development and Recovery," "Assessment of Sustainable Development Indicators and Strategies," "Economic Development and Innovation in Ukraine," and education-related topics like "Digital Transformation in Higher Education" and "Education Reform and Social Responsibility." Agricultural science and climate change adaptation also performed well. This strongly indicates that despite facing immense difficulties from conflict and infrastructure damage, domestic research efforts in Ukraine are refocusing priorities and resources towards addressing urgent national challenges such as economic reconstruction, education system reform, sustainable development, and food security (Ivanchenkov et al., 2024; Shulga, 2023). While topics with negative slopes also exist among the staying group, their absolute values are generally smaller than the largest negative slopes observed in the "departed" group. For example, basic physics fields like "Theoretical and Experimental Nuclear Structure" and "Interactions of Low-Energy Electrons with Matter and Atoms" show negative growth, potentially due to reduced research funding, constrained experimental conditions, damaged or inaccessible equipment. However, the relatively smaller magnitude of decline may suggest that scholars in these fields are endeavoring to maintain minimal research activity, perhaps shifting towards theoretical or computational research to compensate for the lack of experimental facilities.

As shown in Fig. 4, among the group of Russian scientists who left their home country, some basic science fields demonstrated significant resilience and growth momentum, such as "Search for Quark-Gluon Plasma in Heavy-Ion Collisions" and "Particle Physics and High-Energy Collider Experiments." As indicated in Fig. 2, these scholars primarily migrated to scientifically advanced countries like the United States, China, and Germany. These countries are significant research hubs in fields such as high-energy physics, possessing world-leading large-scale experimental facilities and active international collaboration networks. By migrating to these regions, scholars can more easily integrate into and continue to utilize these global resources and collaborative platforms, thereby maintaining or even enhancing their research output. However, this group experienced a steep decline in other fields, particularly certain areas of medicine, genomics, and condensed matter physics and quantum computing. Although these scholars may have also moved to scientifically advanced nations, this downward trend strongly suggests that the continuation of research in these directions is highly constrained by dependence on specific, often Russia-localized resources (such as data from specific patient populations, unique laboratory environments, or equipment). Even in leading research countries, it may be difficult to fully replicate or replace these unique resources originating from Russia, thereby limiting progress in related research. This highlights the "local" nature of research in certain disciplines or a high degree of reliance on specific infrastructure, making it difficult for the physical migration of talent to fully bridge the disruption in research conditions.

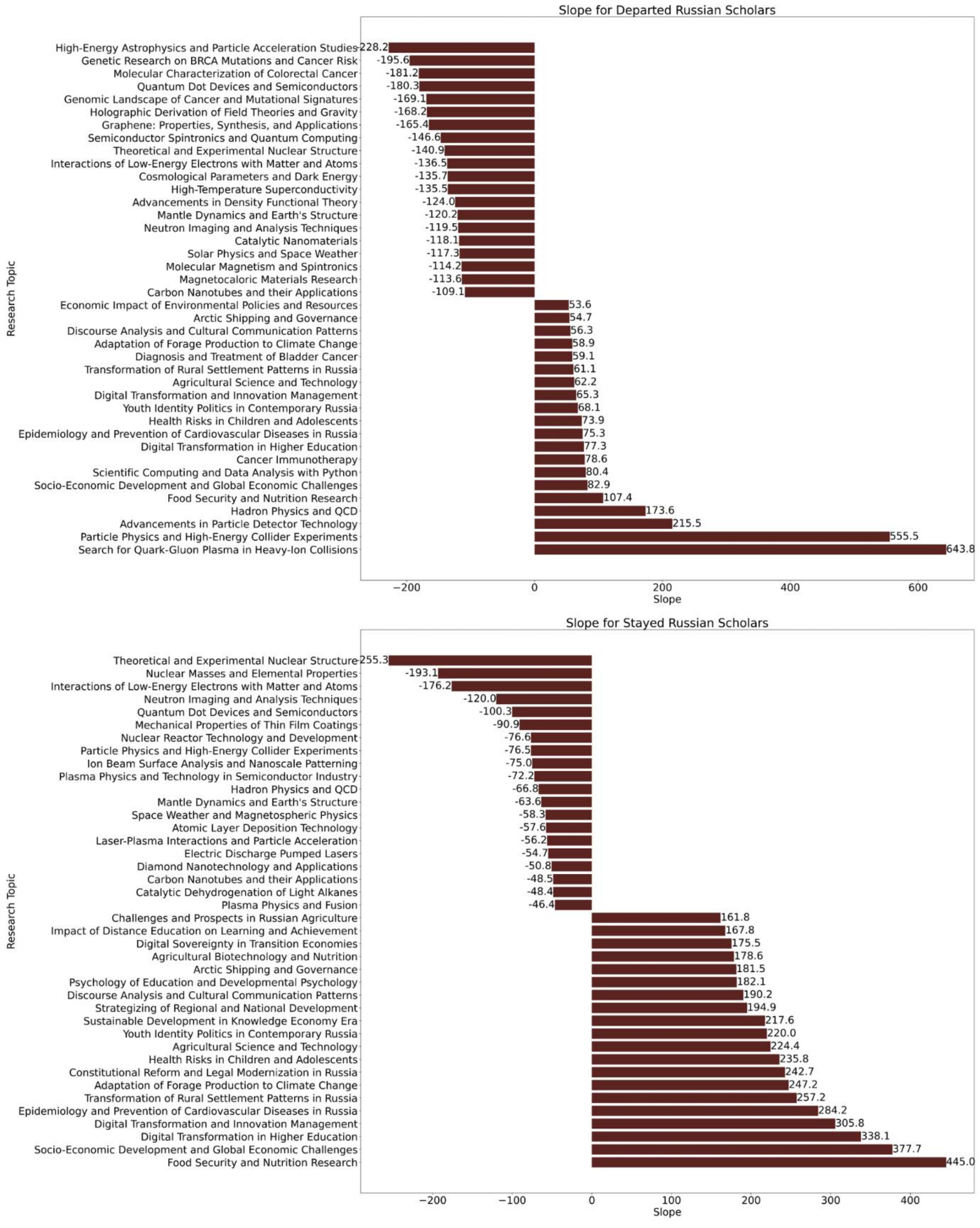

**Figure 4. Trends in research topics over time for Russian Scholars (Departed vs. Stayed).**

In contrast, the group of Russian scientists who remained in their home country shows different priorities and challenges. As depicted in Fig. 4, the areas with the fastest growth in their research output are concentrated on topics closely related to Russia's domestic social, economic, and technological development, such as "Food Security and Nutrition Research," "Socio-Economic Development and Global Economic Challenges," and social science research focusing on domestic issues. This trend is likely driven primarily by the Russian government's readjustment of research priorities in the face of sanctions and technological isolation, channeling resources and policy support towards fields deemed strategically important for national security or achieving "import substitution."(Epifanova, 2022). The shift in research direction among the remaining scientists to serve national development needs, often under resource constraints, constitutes a major line of development within the current Russian research system.

Scholars who remained in Russia experienced significant negative growth in many basic and experimental science fields, particularly nuclear physics, some areas of materials science, and plasma physics. This decline is not simply due to a reduction in personnel but more directly reflects the systemic impact of international sanctions on Russian research infrastructure: restricted access to cutting-edge experimental equipment, key components, reagents, consumables, and software licenses; a sharp reduction in opportunities for international academic exchange and collaboration; and even increased difficulty in accessing the latest research literature and publishing findings in high-impact international journals. These factors collectively pose severe challenges to Russian basic research and experimental disciplines, leading to a significant downward trend in their output. The fact that some fields show negative growth in both the departed and staying scholar groups further highlights potential structural or policy-related deep-seated difficulties facing these disciplines across the entire Russian research system, regardless of where the scholars are located. Since 2014, geopolitical tensions between Russia and Western countries have continuously escalated, peaking in 2022. This series of events and the severe international sanctions they triggered form the core context for understanding the current landscape of the Russian scientific community (European Affairs Committee, 2024). Combined with the main talent destinations shown in Fig. 1, we observe that the conflict not only caused talent outflow but, more importantly, this outflow was not random. It directed talent towards major global research centers. This allowed some scholars to continue their frontier research in new environments but also exacerbated "talent bleed" and research disruption in specific fields within Ukraine and Russia. Simultaneously, the research capacity remaining within the home countries, operating under resource constraints or international isolation, has been compelled to reorient priorities towards serving national strategic needs, leading to significant damage to many disciplines dependent on international exchange and resources. Figures 3 and 4, combined with Fig. 2, powerfully illustrate how geopolitical conflict, mediated by talent flow and its destinations, jointly shapes the long-term evolution of a country's scientific innovation capacity, talent base, and international academic standing.

### 4.3 Impact of Geopolitical Tensions on International Academic Collaboration

To explore the specific changes in collaboration patterns closely, we examined the annual collaboration activity of Ukrainian and Russian scholars with their top 20 most frequent collaborating countries, as illustrated in Figure 5a for Ukrainian cohorts and Figure 5b for Russian cohorts. For Ukrainian scholars remaining in the country (Figure 5a, UA Stayed Active Scholars), the data suggests a tendency toward closer integration with academic networks in several European countries. Within this group, partnerships with countries in Eastern Europe (such as Poland and the Baltic states) showed an upward trend, with their collaboration share rising from 13.90% in 2013 to 18.95% in 2021, and reaching 23.18% in 2023. Concurrently, collaborations with countries in Western and Northern Europe exhibited a complex trajectory; the share was 25.20% in 2013, adjusted to 18.67% in 2021, and subsequently reached 30.57% in 2023. These adjustments may be linked to both individual scholarly initiatives and institutional support programs established within the European Research Area (Plackett, 2022). Migrated Ukrainian scholars (Figure 5a, UA departed) also maintained relatively stable collaborative ties with countries in Northern America, which accounted for 8.82% in 2013 and 9.23% in 2023, while their engagement with Western European partners was recorded at 26.03% in 2023. Ukraine’s full association with the Horizon Europe program provided critical financial incentives that formalized and accelerated these international collaborative trajectories (European Commission, 2024).

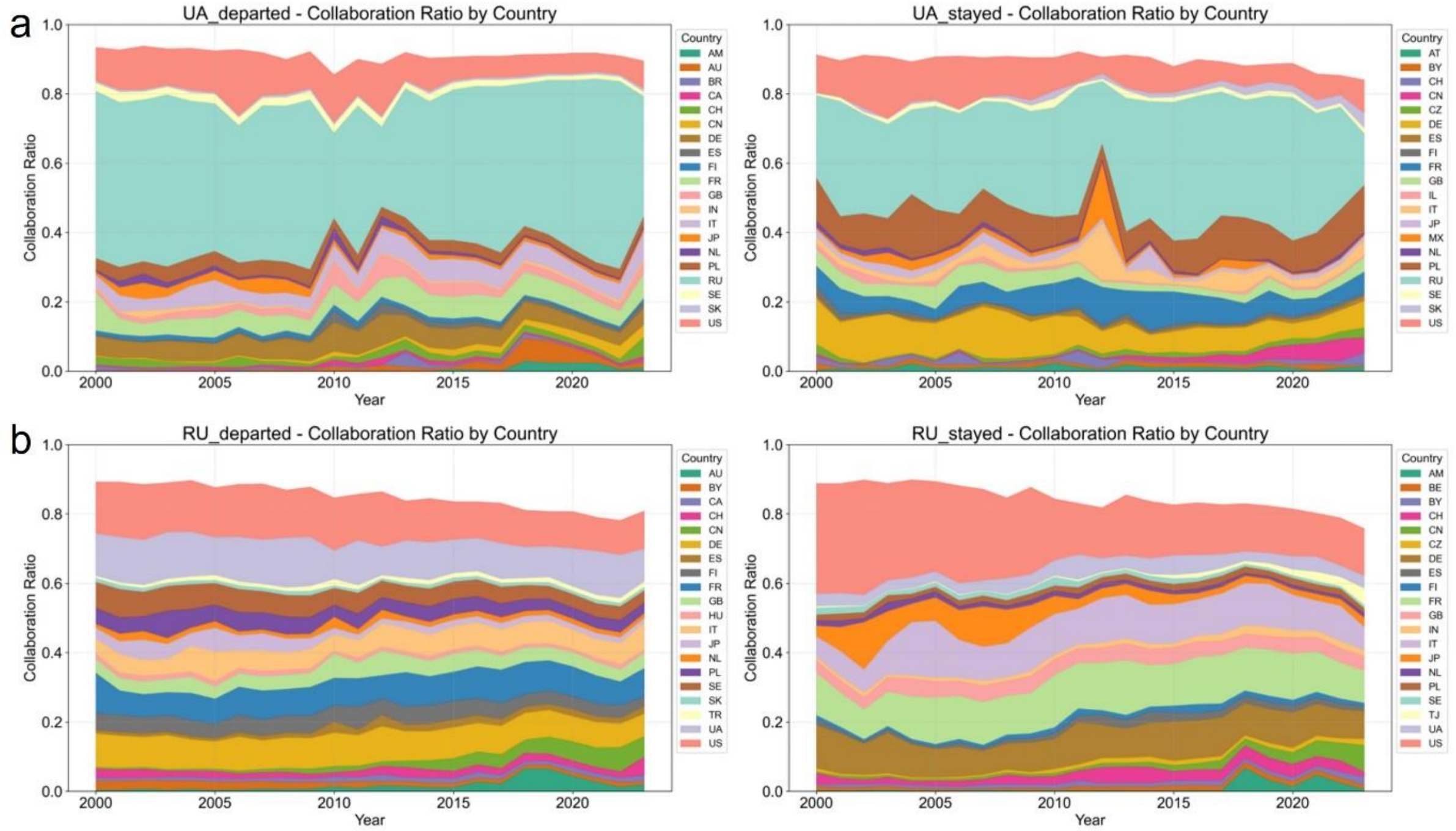

**Figure 5: Impact of Geopolitics on International Academic Collaboration of Russian and Ukrainian Scholars.** (a) Changes in Major International Collaboration Partners for Ukrainian Scholars (Departed vs. Stayed). (b) Changes in Major International Collaboration Partners for Russian Scholars (Departed vs. Stayed).

For Russian scholars, the longitudinal data in Figure 5b indicates a gradual reconfiguration of international ties, characterized by a reduction in partnerships with several Western nations and an expansion toward other regions. Among scholars remaining in Russia (Figure 5b, RU Stayed Active Scholars), collaborative ties with countries in Western and Northern Europe contracted from 40.65% in 2013 to 34.99% in 2021, and were measured at 29.99% in 2023. A similar trend appears in partnerships with Northern American countries, which changed from 18.35% in 2013 to 13.26% in 2021. This reduction in traditional collaborative channels may reflect the broader institutional and geopolitical environment (Plackett, 2022). Simultaneously, the data shows an increase in the share of collaborations with countries in Eastern and SE Asia (primarily China), which rose from 5.46% in 2013 to 12.62% in 2023. Following the escalation and subsequent sanctions, Russian policy incentives have undergone a strategic reorientation. The focus has shifted from Western-centric integration to the "Priority 2030" program, which emphasizes technological sovereignty while incentivizing new collaborative axes with BRICS and Shanghai Cooperation Organization (SCO) nations (MSHE, 2021).

Further observation of the Russian results (Figure 5b) suggests a diversification of collaborative networks toward other regional hubs. Partnerships with countries in West and Central Asia (such as Kazakhstan and Uzbekistan) increased from 2.36% in 2013 to 6.60% in 2023. Additionally, the share of collaborations with countries in the "Rest of the World" category—representing various emerging scientific powers—increased from 4.23% in 2013 to 11.74% in 2023. For migrated Russian scholars (Figure 5b, RU departed), while they maintained a higher baseline of collaboration with Western and Northern European countries (35.79% in 2023) compared to their counterparts who stayed, they also showed a distributed engagement with Southern European countries and other global nodes (Mallapaty et al., 2022). These multidisciplinary findings suggest that research networks are undergoing a process of structural adjustment, moving toward a more diversified global landscape where collaborative alliances appear to be increasingly influenced by the broader geopolitical context.

While these overarching share-based trends illustrate a macro-level geopolitical realignment, the structural changes appear to be distributed heterogeneously across scientific fields and scholar cohorts. To further explore these dynamics, **Supplementary Note 9** provides a granular longitudinal analysis across 26 core disciplines, alongside an observation of spatial reconfiguration based on absolute-count metrics. This analysis suggests significant disciplinary variation and a notable divergence between different scholar groups. For instance, certain Russian applied sciences exhibit a trend toward an "eastward shift" in collaborative ties, whereas social sciences show signs of a more fragmented network. Interestingly, these shifts are more pronounced among scholars remaining in Russia, while those who migrated appear to maintain a collaborative profile more closely aligned with global mainstream networks. For the Ukrainian research community, the data reflects a systematic realignment away from legacy post-Soviet ecosystems, establishing closer ties with Western and Central-Eastern European partners. This transition is characterized by the formation of "proximity sanctuaries" with nations such as Poland and a deepened integration with Western European academic networks—a trend observed among both stayed and departed Ukrainian cohorts. Furthermore, regional observations suggest that while domestic research systems are developing new regional anchors (such as Kazakhstan and Uzbekistan for the Russian system), individual mobility and host-country science policies for scholars at risk remain key factors in navigating institutional barriers.

### 4.4 Academic Role Changes and Challenges for Migrated Scholars in Host Countries

As shown in Fig. 6, when integrating into new academic environments and conducting research, the trajectory of academic role changes for departed Ukrainian scholars exhibits significant heterogeneity. Despite a general trend towards shifting from initial participation (higher proportion of middle authorship) towards leading positions (increasing proportion of first and corresponding authorship), the pace and pattern of this shift vary significantly across different receiving countries, reflecting the differing influences of host country academic systems, support policies, and integration environments.

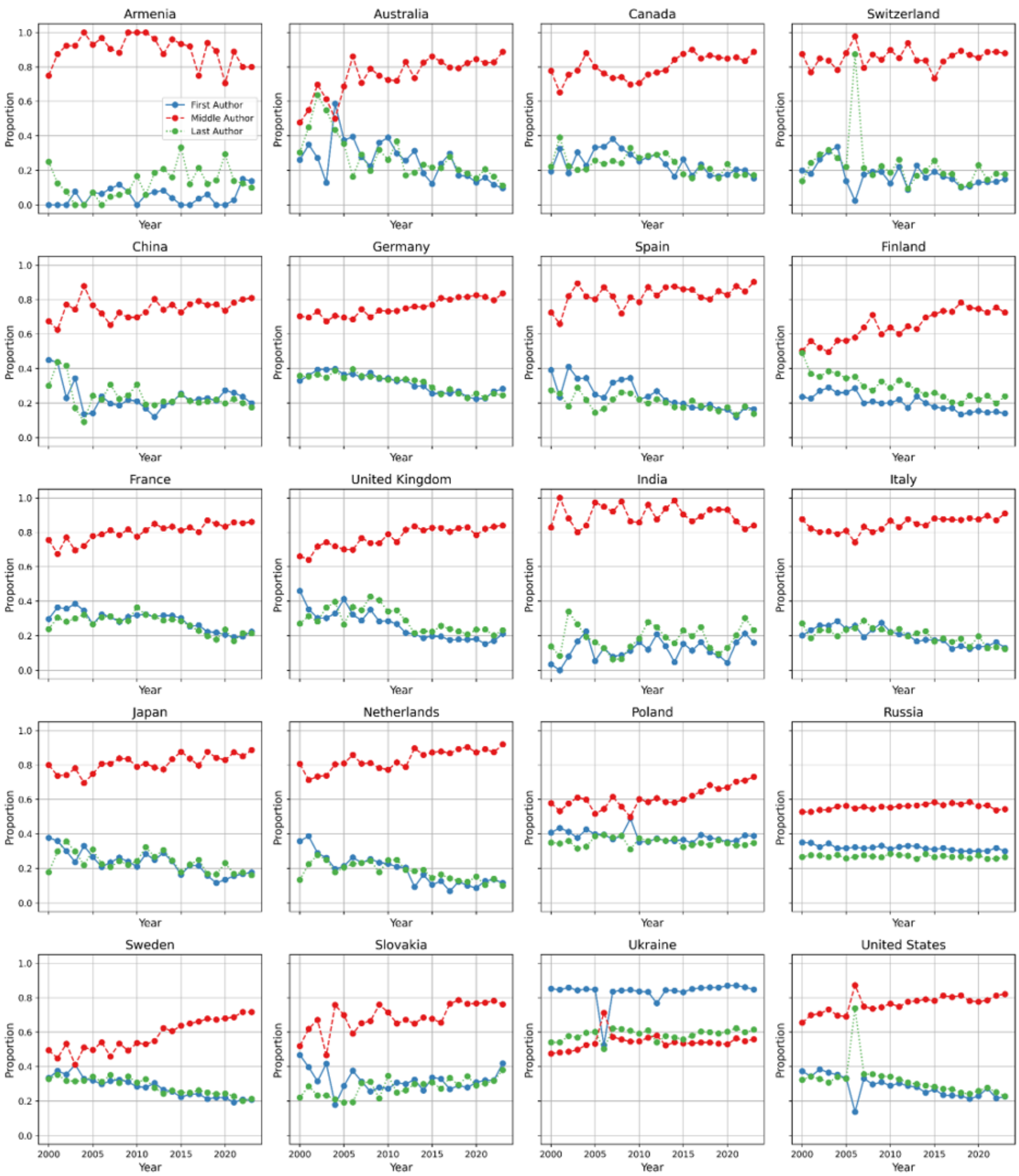

**Figure 6. Academic role trajectories (authorship positions) of departed Ukrainian scholars in major receiving countries.**

In the vast majority of major receiving countries, including Australia, Canada, Germany, Finland, the United Kingdom, France, Japan, the Netherlands, Poland, Sweden, and the United

States, a notable trend is generally observed: the proportion of research where Ukrainian scholars are listed as middle authors shows a substantial annual increase, potentially rising by 10% to 20%. While this significant and continuous increase in the proportion of middle authorship superficially suggests that Ukrainian scholars are actively participating in research activities in these countries and successfully integrating into existing research teams, contributing their expertise, this pattern may conceal the arduous challenges and structural barriers faced by departed scholars. In many cases, a middle authorship position signifies participation and contribution but typically does not imply being the project leader, concept originator, or head of the research direction. This shift in the center of gravity of authorship positions may reflect difficulties in quickly obtaining independent research positions or establishing their own research teams in their new environments. However, this phenomenon is not merely a sign of successful integration in a simple sense, but rather more likely represents an adaptation strategy for a soft landing in difficult circumstances. Facing the immense pressure of research interruption, dissolution of former teams, broken funding chains, and starting anew in a different country, joining existing, often larger, research teams and participating in publications as middle authors allows them to maintain research activity during a challenging period, ensuring a degree of academic output and preventing their careers from completely stalling.

The high proportion of middle authorship is also likely reinforced by the nature of emergency policy responses in host countries. Initiatives such as MSCA4Ukraine and various national fellowships in Germany and the UK often focus on providing immediate "hosting" opportunities within established research groups (Alexander von Humboldt Foundation, 2023; Scholars at Risk Europe, 2023). While these policies are vital for scholar preservation, they structurally position displaced researchers as collaborators within existing institutional hierarchies rather than as independent principal investigators, which is reflected in their authorship roles.

In contrast to the prevalent pattern described above, the sub-figures for a few receiving countries (such as Armenia and Slovenia) show a smoother integration pattern. After Ukrainian scholars arrived in these countries, their proportion of first authorship and corresponding authorship (typically the last author) increased relatively quickly and remained at a high level in subsequent years. This pattern suggests that in Armenia and Slovenia, Ukrainian scholars were able to transition relatively quickly from being collaborators to becoming leaders and principal investigators of research projects. The relatively smaller scale of the academic communities in Armenia and Slovenia may offer a context with fewer structural barriers (see **Supplementary Note 10**). This environment potentially facilitates the identification and integration of experienced migrated scholars into research teams, and might create more accessible pathways toward independent research resources or leadership roles within a shorter timeframe, compared to the more complex hierarchies of larger scientific systems. Among the group of Ukrainian scholars who migrated to Switzerland, China, Spain, and India, the distribution of their authorship roles in publications showed relative stability over time, with no significant large changes observed. The proportions of first authors, middle authors, and corresponding authors remained roughly constant.

Compared to the complex and varied changes in research roles for Ukrainian scholars abroad, as shown in Fig. 7, the authorship role changes exhibited by Russian departed scholars in host countries demonstrate a high degree of consistency and a significant, negative pattern biased towards passive participation. Notably, among the group of Russian scholars who migrated to Switzerland, China, Spain, and India, the distribution of their authorship roles in publications showed relative stability over time, with no significant large changes observed. However, in almost all other countries where they frequently migrated, the proportion of publications where Russian scholars held leading research roles (first and corresponding authorship) systematically experienced a substantial decrease, while the proportion of participation as middle authors generally increased significantly. This pattern stands in stark contrast to the cases where some Ukrainian scholars successfully transitioned to leading roles in certain countries. Even in Hungary and Turkey, while there was a slight increase in the proportion of Russian scholars holding leading research roles, the proportion of middle authors still showed a growing trend, not altering the fundamental pattern of becoming collaborators rather than leaders overall. Figure 7, which reveals the widespread phenomenon of Russian departed scholars becoming predominantly middle authors, strongly suggests that after migrating to new environments, they face a common and difficult situation in regaining or establishing academic leadership commensurate with their previous senior status. This pattern may be related to the lack of specific support mechanisms for Russian scholars in host countries, the influence of political factors and international sanctions on receiving institutions' decisions, and structural barriers preventing Russian scholars from securing leading roles in the new system, resulting in a long-term negative impact on their academic leadership potential and independent research capacity (Mäkinen, 2023).

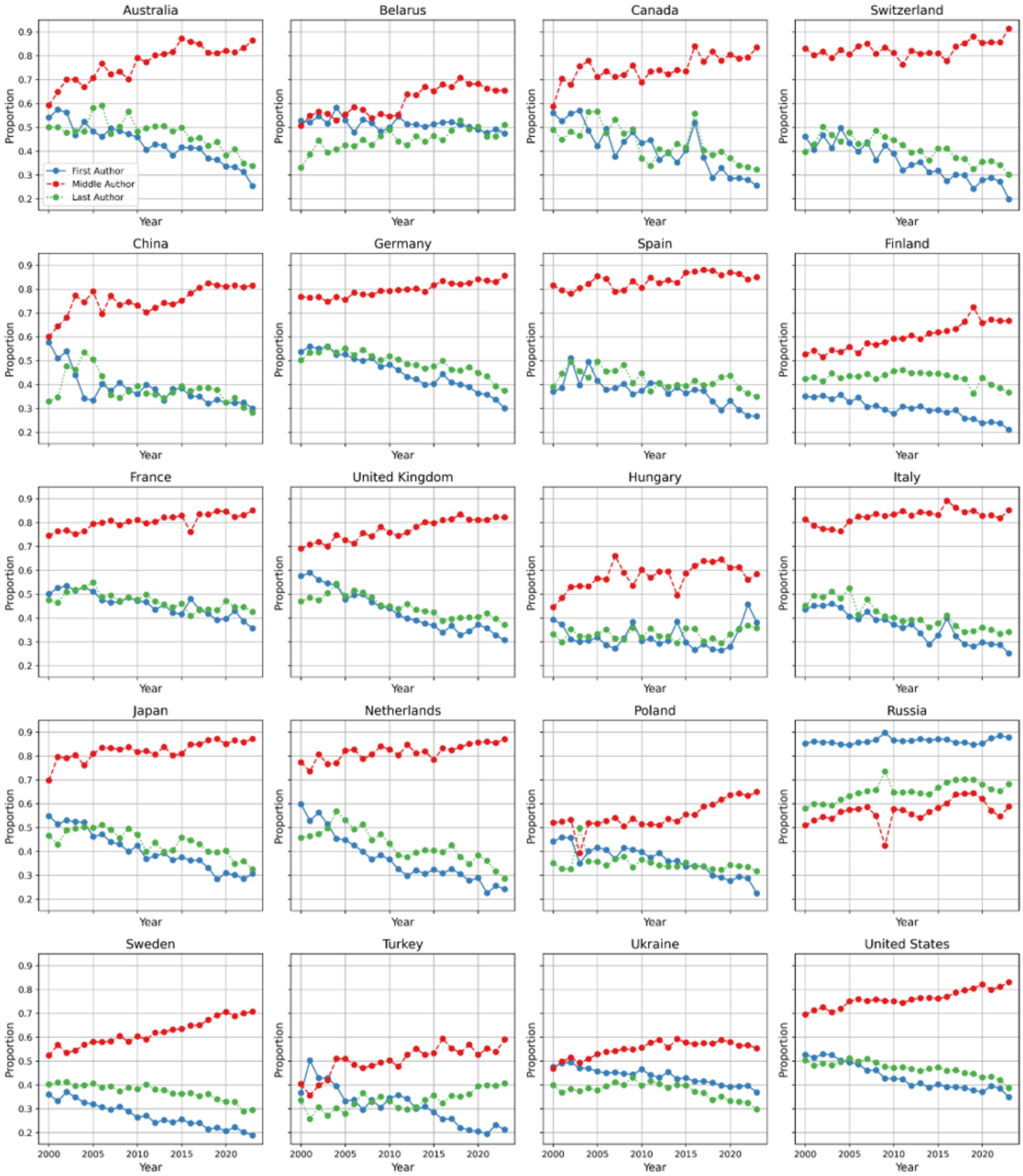

**Figure 7. Patterns of authorship role changes for departed Russian scholars in major receiving countries.**

To further evaluate the robustness of these observations, a detailed analysis was conducted across ten major research fields as presented in **Supplementary Note 11**. The results suggest that scholars remaining in their home countries consistently maintain a pattern of academic leadership, with a high proportion of first authorship in fields such as Engineering, Chemistry, and Materials Science. In contrast, for migrated cohorts, a shift toward increased middle authorship is observed as a common trend across various disciplines. This transition is particularly evident in collaborative fields such as Medicine and Biochemistry, yet a comparable increase in middle-authorship roles was also found in areas with different publication norms, including the Social Sciences. The consistency of these patterns across several distinct disciplines indicates that the observed change from independent leadership to collaborative participation is associated with the structural adjustments following migration rather than being a reflection of specific disciplinary authorship conventions.

# 5 Discussion

Our quantitative analysis based on comprehensive OpenAlex metadata (2000-2023) provides empirical evidence of the profound and complex impact of escalating geopolitical conflict on the academic ecosystems of Ukraine and Russia. We documented a sharp acceleration in scholar brain drain, a fundamental reshaping of research priorities and output, significant alterations to international collaboration networks, and critical challenges faced by displaced scholars in establishing themselves in new environments.

The conflict, particularly the 2022 escalation, triggered unprecedented academic talent migration from both countries. Our analysis quantified these effects, showing that while the 2014 events initiated a gradual upward trend in scholar departures from both Ukraine and Russia, the 2022 full-scale invasion resulted in an immediate, massive surge in scholar outflow, followed by a sustained, accelerating rate of departure. Despite Russia's significantly larger scholar population pre-war compared to Ukraine, the magnitude of the long-term outflow trend after 2022 was strikingly similar for both countries. This implies a disproportionately devastating impact on Ukraine's smaller research base, potentially leading to a severe depletion of its academic workforce. While Russia may see some mitigation through ties with neighboring countries and potential returnees, the sheer scale of the outflow represents a significant brain drain with serious long-term consequences for the national scientific capacity of both nations.

Furthermore, the conflict led to a clear divergence in the research landscape between departed and stayed active scholars, as well as between the two countries. Departed Ukrainian scholars showed resilience and growth in highly internationalized basic science fields. However, they experienced sharp declines in fields heavily reliant on specific, often location-bound resources, such as certain medical specialties and experimental physics areas. Conversely, scholars remaining in Ukraine demonstrated remarkable adaptation under extreme adversity, strategically shifting towards applied research directly relevant to national reconstruction and societal needs, although basic experimental fields also showed some decline, albeit less severe than for departed scholars. These findings corroborate recent evidence that infrastructure-heavy institutions, such as NASU, have faced more acute collaboration disruptions than universities (Hladchenko, 2026). Our study extends this by revealing how such structural vulnerabilities specifically trigger a strategic reorientation toward applied research areas. For Russia, departed scholars similarly thrived in international basic science centers but experienced substantial shrinkage in fields tied to Russia-specific resources (e.g., patient data, unique facilities). Crucially, stayed Russian scholars, operating under international sanctions and political isolation pressures, focused on nationally prioritized strategic areas (e.g., food security, socio-economic development). However, this reorientation was accompanied by a significant decline in many basic and experimental science fields, directly attributable to systemic barriers in accessing international resources, equipment, collaboration channels, and publication venues. These diverging trends highlight how geopolitical pressure not only drives talent migration but also leads to fragmentation of research efforts and compels strategic reorientation, potentially at the expense of breadth and foundational research.

The conflict also reshaped international academic collaboration patterns. Post-2014, there was a significant decoupling of academic ties between Ukraine and Russia, underscoring how geopolitical faultlines override historical connections and geographical proximity. While Ukrainian scholars (both departed and stayed) increasingly sought collaborations with Western countries, reflecting solidarity and a repositioning within global networks, Russian scholars saw a significant decline in collaboration with traditional Western partners, particularly after 2022. This decrease was partially offset by increasing collaboration with China, indicating a strategic pivot towards non-Western academic partnerships. These shifts risk marginalizing academic communities in both countries from certain international research streams and knowledge diffusion channels, while potentially establishing new axes of collaboration.

Our analysis of authorship positions reveals significant challenges faced by departed scholars in establishing themselves in new academic environments, particularly for those from Russia. While Ukrainian scholars showed varied patterns, with some integrating relatively quickly into leading roles (first/last author) in certain receiving countries (e.g., Armenia, Slovenia), a general trend in most major receiving countries was a substantial increase in their proportion of middle authorship. This indicates active participation in research but may also reflect difficulties in securing independent positions or leading research projects, perhaps representing an adaptive strategy to maintain productivity. For Russian scholars, however, this pattern was highly consistent and negative across almost all major receiving countries, showing a widespread significant decline in first and last authorship positions and a rise in middle authorship proportion. This contrasts sharply with some of the more positive experiences of some Ukrainian scholars and points to potential structural barriers, lack of targeted support, and perhaps the indirect impact of political factors and sanctions on receiving institutions' decisions, all likely affecting their ability to achieve independent research leadership within host institutions.

This study highlights the disruptive consequences of geopolitical conflict on critical components of the scientific ecosystem, including human capital, thematic orientations, international linkages, and individual academic careers. The large-scale displacement of scholars, the fragmentation of disciplinary focus, and the difficulties faced by displaced researchers in re-establishing their academic footing pose significant long-term threats to the research capacities of both Ukraine and Russia. Moreover, the ongoing reconfiguration of international collaborations risks reshaping the global science landscape, potentially leading to the erosion of domain-specific knowledge and a weakening of cumulative knowledge production. Importantly, these developments extend beyond national boundaries. They represent a structural challenge to the coherence and resilience of global science. Policy responses should not only support displaced scholars through emergency fellowships but also prioritize long-term frameworks for reintegration and capacity rebuilding. Programs such as Harvard University's Scholars Without Borders offer a promising model—focusing on supporting displaced scholars from Ukraine, Russia, and Belarus to continue their academic work despite the disruptions caused by war and political repression, rather than drawing them into competitive talent migrations that may hinder long-term reconstruction in their home countries (Huang & Yu, 2024).

This study is not without limitations. Given academic publishing lags, our 2023 cutoff captures the acute systemic shock rather than long-term structural adjustments. Future research utilizing 2024+ datasets is necessary to observe stabilized post-migration output and determine if these initial disruptions settle into a new equilibrium. Regarding data coverage, we acknowledge the inherent metadata limitations of automated scholarly databases such as OpenAlex (Zhang et al., 2024), acknowledging that no bibliographic infrastructure is entirely without imperfections. These constraints imply that the study may undercount scholars who are primarily active in regional or non-indexed publication venues. Furthermore, we applied stringent classification criteria to distinguish between stayed active scholars and migrated scholars, resulting in a highly conservative lower-bound estimation of talent mobility. While this methodology prevents the artificial inflation of brain drain figures, it may overlook individuals who have physically relocated yet continue to maintain formal institutional affiliations. Furthermore, the results should not be interpreted as definitive causal effects.

## Data availability

Due to licensing restrictions, the raw data from OpenAlex used in this study cannot be shared. However, the processed datasets generated and analyzed during the research are available in anonymized form via the Open Science Framework at https://doi.org/10.17605/OSF.IO/8SZE6. To protect the privacy of individual scholars and prevent potential risks to their safety arising from the ongoing conflict, all personally identifiable information has been removed. The shared data include anonymized migration records of Ukrainian and Russian scholars, containing non-identifiable scholar IDs and institution IDs, along with the actual geographic coordinates of institutions. Additionally, we provide annual data on the number of migrated and non-migrated scholars from Ukraine and Russia, their destination countries, and corresponding coordinates. We also include data of 2,000 research topics—those with the largest increases and decreases in output—for both migrated and non-migrated scholars in each country, along with detailed topic-level information. Furthermore, the dataset contains yearly collaboration records between Ukraine, Russia, and other countries, as well as authorship position data across different scientific publications.

## Competing interests

The authors declare no competing interests.

## Ethical approval

This study uses publicly available bibliometric data from OpenAlex. No human participants were involved, and no personally identifiable information was collected or analyzed. All data were anonymized prior to analysis to protect the privacy and safety of individuals.

## Informed consent

This article does not contain any studies with human participants or their data.

# Supplementary Notes for **Frontlines and faultlines: How the Russo-Ukrainian conflict reshapes the landscape of scientific research**

# SUPPLEMENTARY NOTE 1: TECHNICAL DISCLOSURE OF METADATA MISSINGNESS IN THE OPENALEX DATABASE (2000–2023)

**1. Introduction to Metadata Harvesting and Attribution Constraints**

The OpenAlex database operates as a large-scale, open-access aggregator of scholarly metadata, utilizing automated harvesting and algorithmic disambiguation. As an endogenous characteristic of such systems, there exists a significant proportion of records where institutional and geographic identifiers are not successfully resolved. This Note provides a multi-dimensional disclosure of these missingness patterns across the 2000–2023 observation window to establish the empirical boundaries of the data utilized in this study.

**2. Macro-Level Work Metadata Missingness: Global Baseline (2000–2023)**

The most direct measure of metadata missingness is the work-level country code availability. We define "No Country Info" as instances where a publication record fails to link any of its authors to a standardized country code. Table S1 presents the comprehensive annual distribution of this missingness across 187 million records.

**Table S1. Annual distribution of global work-level geographic metadata missingness.**

| Year | Total Works Analyzed | Works with No Country Info | Missing Rate (%) |
|---|---|---|---|
| 2000 | 3,531,505 | 2,262,382 | 64.06 |
| 2001 | 3,713,479 | 2,378,288 | 64.04 |
| 2002 | 4,317,823 | 2,602,495 | 60.27 |
| 2003 | 4,558,926 | 2,827,071 | 62.01 |
| 2004 | 4,975,956 | 3,118,236 | 62.67 |
| 2005 | 5,342,204 | 3,262,598 | 61.07 |
| 2006 | 5,829,746 | 3,490,384 | 59.87 |
| 2007 | 6,363,744 | 3,876,271 | 60.91 |
| 2008 | 6,742,869 | 4,041,390 | 59.94 |
| 2009 | 7,302,007 | 4,397,539 | 60.22 |
| 2010 | 7,813,240 | 4,704,432 | 60.21 |
| 2011 | 8,494,128 | 5,123,738 | 60.32 |
| 2012 | 8,748,299 | 5,264,730 | 60.18 |
| 2013 | 9,213,114 | 5,526,397 | 59.98 |
| 2014 | 9,716,265 | 5,890,707 | 60.63 |
| 2015 | 9,862,699 | 5,975,894 | 60.59 |
| 2016 | 10,200,404 | 6,244,538 | 61.22 |
| 2017 | 10,038,483 | 5,845,346 | 58.23 |
| 2018 | 10,090,787 | 5,537,508 | 54.88 |
| 2019 | 10,375,921 | 5,483,974 | 52.85 |
| 2020 | 11,066,876 | 5,510,881 | 49.80 |
| 2021 | 9,869,644 | 4,155,419 | 42.10 |
| 2022 | 9,479,974 | 3,363,463 | 35.48 |
| 2023 | 9,922,200 | 4,167,742 | 42.00 |
| Overall | 187,570,293 | 105,051,423 | 56.01 |

## 3. Institutional Entity Missingness: Knowledge Base Integrity

While work-level missingness involves the failure of algorithmic linking, institution-level missingness refers to the absence of geographic attributes within the database's core entity registry. Table S2 discloses the metadata status of all unique institutions currently indexed in the OpenAlex Knowledge Base.

**Table S2. Disclosure of geographic metadata in the institutional entity registry.**

| Entity Status | Absolute Count | Percentage (%) |
|---|---|---|
| Total Unique Institutions (Full Registry) | 114,883 | 100.00 |
| Institutions with Country Code Assigned | 102,467 | 89.19 |
| Institutions Missing Country Code | 12,416 | 10.81 |

## 4. Annual Missingness Trends in Active Research Infrastructure

To disclose the dynamic quality of the institutional knowledge base, we audited the "Active Institutions" (those appearing in at least one publication) for each year from 2000 to 2023, as shown in Table S3. This measures the database's ability to geocode the entities participating in global research in real-time.

**Table S3. Annual geographic metadata missingness for active institutions.**

| Year | Active Institutions | Missing Country Code | Missing Rate (%) |
|---|---|---|---|
| 2000 | 43,864 | 1,106 | 2.52 |
| 2001 | 45,358 | 1,203 | 2.65 |
| 2002 | 48,223 | 1,302 | 2.70 |
| 2003 | 49,794 | 1,414 | 2.84 |
| 2004 | 51,373 | 1,482 | 2.88 |
| 2005 | 53,542 | 1,682 | 3.14 |
| 2006 | 55,178 | 1,802 | 3.27 |
| 2007 | 56,847 | 1,929 | 3.39 |
| 2008 | 58,565 | 2,093 | 3.57 |
| 2009 | 60,260 | 2,202 | 3.65 |
| 2010 | 61,833 | 2,357 | 3.81 |
| 2011 | 63,613 | 2,601 | 4.09 |
| 2012 | 65,240 | 2,757 | 4.23 |
| 2013 | 66,646 | 2,916 | 4.38 |
| 2014 | 67,840 | 3,115 | 4.59 |
| 2015 | 69,166 | 3,321 | 4.80 |
| 2016 | 69,764 | 3,502 | 5.02 |
| 2017 | 70,380 | 3,671 | 5.22 |
| 2018 | 71,094 | 3,918 | 5.51 |
| 2019 | 71,824 | 4,205 | 5.85 |
| 2020 | 72,936 | 4,530 | 6.21 |
| 2021 | 72,814 | 4,699 | 6.45 |
| 2022 | 72,302 | 4,823 | 6.67 |
| 2023 | 72,243 | 4,970 | 6.88 |

## 5. Summary of Global Metadata Limitations

The multi-level disclosure presented in this Note establishes the technical boundaries of the OpenAlex metadata for the period 2000–2023. The data reveals a significant linkage gap within the database architecture. While 56.01% of individual work-level records lack standardized country-level attribution, the underlying Institutional Registry demonstrates a much higher geographic coverage, with 89.19% of all unique entities successfully geocoded. This suggests that the primary challenge in utilizing OpenAlex for migration studies is the algorithmic failure to link raw publication strings to known institutional identifiers rather than a fundamental absence of geographic knowledge within the Knowledge Base.

# SUPPLEMENTARY NOTE 2: INTERNAL METADATA AUDIT AND CLASSIFICATION LOGIC FOR RUSSIAN AND UKRAINIAN SCHOLARS

## 1. Internal Metadata Integrity of the RU/UA Target Populations

Before evaluating migration trends, it is necessary to disclose the metadata quality specifically within the identified Russian and Ukrainian (RU/UA) scholarly populations. It is important to clarify that this audit pertains exclusively to the visible cohort, consisting of those scholars whose records contained sufficient information to be assigned a country code during the identification process. The audit demonstrates that within the identified RU/UA corpus, affiliation density is relatively high, with approximately 85% to 87% of works possessing a complete set of author affiliations, as shown in Table S4. The overall internal missing rate for these populations is observed to be approximately 4.8%, which is significantly lower than the global work-level baseline. This high internal fidelity suggests that for the specific cohort captured in this study, longitudinal trajectories possess a degree of structural continuity that is less susceptible to fragmentation than the broader database average.

**Table S4: Internal metadata quality audit report for RU/UA populations (2000–2023)**

| Statistical Metric | Russia | Ukraine |
|---|---|---|
| Total Works Analyzed | 2,437,583 | 520,587 |
| Total Author-Paper Links | 20,092,115 | 5,394,329 |
| Missing Affiliation Identifiers | 983,068 | 256,815 |
| Overall Internal Missing Rate (%) | 4.89 | 4.76 |
| Average Missing Authors per Work | 0.4033 | 0.4933 |
| Perfect Metadata Works (Zero Missing) (%) | 87.45 | 85.06 |
| Ghost Works (All Affiliations Missing) (%) | 0.00 | 0.00 |

## 2. Disclosure of the Invisible Population and Journal-level Metadata Attrition

Despite the high internal fidelity shown in Table S4, the study must cautiously acknowledge that these figures do not imply absolute completeness of the entire scholarly population. As disclosed in Supplementary Note 1, the OpenAlex global baseline exhibits a 56.01% institution missing rate at the work-level. This macro-level attrition suggests that a substantial volume of publications, particularly those in regional, non-indexed, or historically non-digitized venues, may fail to provide structured affiliation strings in the raw metadata. In such instances, even if the Institutional Registry is robust, the identification algorithm lacks the necessary input signal to link a paper to a specific country. This implies the existence of an invisible population of Russian and Ukrainian scholars whose publication history may lack traceable identifiers across the entire observation window. Consequently, the migration dynamics captured here should be interpreted as a conservative estimation focusing on the professionalized core of the RU/UA scientific workforce who are predominantly active within more standardized scholarly niches.

**3. Institutional Registry and Structural Visibility**

The study partially relies on the Institutional Registry, which demonstrates a 89.19% global geographic coverage, as a supportive framework. However, this registry serves only as a secondary infrastructure rather than an absolute solution to metadata gaps. The geographic information in the registry can only be activated if the raw text of a publication contains institutional names distinct enough to trigger an algorithmic match. Scholars affiliated with large, prestigious, or well-established research organizations likely have a higher probability of being geocoded due to the relative standardization of their institutional names across various journals. Therefore, the observed trends are likely more representative of scholars within the formal, institutionalized research infrastructure, while those at smaller or emerging organizations may be underrepresented in the data.

The low internal missing rate within the identified target sample is a critical factor for the observation of longitudinal trajectories. A high degree of metadata fragmentation would lead to pseudo-disappearances of scholars, which could distort the analysis of their presence over time. The observed 4.8% internal missingness ensures that for the population within the sample, the continuity of their institutional records is sufficient to support a reliable analysis of their geographical presence. This consistency suggests that the observed institutional changes are driven by genuine transitions recorded in the database rather than by random gaps in metadata harvesting.

**4. Cautious Assessment of Statistical Trend Robustness**

While acknowledging the risks of data omission, the study maintains that the structural trends identified by the Interrupted Time Series analysis remain statistically informative. The systematic bias introduced by missing data is expected to be relatively consistent across the temporal axis, provided that the metadata harvesting capabilities of the database did not undergo a non-random or dramatic collapse at the exact moments of the 2014 or 2022 shocks. While the absolute number of scholars may be underestimated, the structural changes, such as shifts in levels and changes in slopes, reflect the community's response to external shocks. These structural markers are generally considered more robust than absolute volumetric counts in bibliometric analysis when dealing with incomplete data.

This disclosure transparently presents the limitations and the technical boundaries of the data. We acknowledge that the 56.01% global missing rate is a significant constraint and that the study may not capture individuals operating exclusively in peripheral publication habitats. However, by demonstrating an internal missing rate of less than 5% for the identified cohort, the study establishes a reliable window into the professionalized segment of the Russian and Ukrainian scientific workforce. The migration dynamics presented herein are offered as a representative, albeit conservative, view of the mobility of core scientific talent from 2000 to 2023.

# SUPPLEMENTARY NOTE 3: QUANTITATIVE ASSESSMENT OF MULTIPLE AFFILIATIONS AND INTERNATIONAL CO-APPOINTMENTS

### 1. Rationale for Evaluating Co-affiliation Trends

In contemporary bibliometric analysis, tracking geographic mobility is inherently complicated by the increasing prevalence of multiple affiliations. Scholars frequently engage in joint appointments, visiting positions, or cross-border collaborations, leading to the concurrent listing of several institutions on a single publication. To transparently address potential biases where temporary international visits might be misclassified as permanent talent drain, this section provides a detailed quantitative assessment of multiple affiliations within the Russian (RU) and Ukrainian (UA) target populations from 2000 to 2023.

### 2. Annual Distribution of Multiple Affiliations

To delineate the scope of this metadata complexity, we extracted longitudinal metrics for each year, including the total number of active scholars, the total number of scholars who reported more than one institutional affiliation on a single paper, and a specific subset of these scholars whose multiple affiliations explicitly included both a domestic institution (RU or UA) and at least one foreign institution. The absolute counts for these metrics are presented in Table S5 (for Russia) and Table S6 (for Ukraine).

**Table S5. Annual trends of multiple affiliations for Russian (RU) scholars (2000–2023).**

| Year | Total Active Scholars | Scholars with Multiple Affiliations | Scholars with Domestic and Foreign Affiliations |
|---|---|---|---|
| 2000 | 39,980 | 7,296 | 3,316 |
| 2001 | 44,769 | 8,453 | 3,843 |
| 2002 | 50,228 | 9,320 | 4,223 |
| 2003 | 48,701 | 9,716 | 4,556 |
| 2004 | 47,887 | 9,753 | 4,588 |
| 2005 | 50,800 | 10,281 | 5,035 |
| 2006 | 55,965 | 11,093 | 5,055 |
| 2007 | 60,731 | 12,757 | 5,837 |
| 2008 | 61,615 | 12,383 | 7,104 |
| 2009 | 62,519 | 12,597 | 7,219 |
| 2010 | 67,383 | 12,578 | 6,766 |
| 2011 | 74,824 | 14,323 | 7,853 |
| 2012 | 79,401 | 14,851 | 7,078 |
| 2013 | 91,402 | 17,895 | 8,485 |
| 2014 | 109,405 | 23,806 | 10,237 |
| 2015 | 132,787 | 30,386 | 10,706 |
| 2016 | 149,212 | 37,003 | 12,685 |
| 2017 | 165,861 | 38,324 | 12,609 |
| 2018 | 194,968 | 45,433 | 14,239 |
| 2019 | 235,397 | 56,165 | 18,451 |
| 2020 | 262,642 | 62,583 | 22,012 |
| 2021 | 283,252 | 64,743 | 22,370 |
| 2022 | 268,559 | 61,267 | 20,306 |
| 2023 | 232,389 | 58,304 | 19,477 |

**Table S6. Annual trends of multiple affiliations for Ukrainian (UA) scholars (2000–2023).**

| Year | Total Active Scholars | Scholars with Multiple Affiliations | Scholars with Domestic and Foreign Affiliations |
|---|---|---|---|
| 2000 | 6,858 | 1,702 | 473 |
| 2001 | 7,547 | 1,835 | 463 |
| 2002 | 8,706 | 2,223 | 562 |
| 2003 | 8,898 | 2,238 | 632 |
| 2004 | 9,182 | 2,357 | 669 |
| 2005 | 8,999 | 2,400 | 620 |
| 2006 | 10,525 | 2,664 | 815 |
| 2007 | 10,460 | 2,719 | 720 |
| 2008 | 10,338 | 2,609 | 972 |
| 2009 | 10,818 | 2,682 | 994 |
| 2010 | 11,975 | 2,825 | 1,027 |
| 2011 | 12,745 | 3,018 | 959 |
| 2012 | 16,077 | 3,066 | 996 |
| 2013 | 19,861 | 3,474 | 1,091 |
| 2014 | 21,663 | 4,019 | 1,300 |
| 2015 | 24,139 | 4,738 | 1,544 |
| 2016 | 30,172 | 5,783 | 1,845 |
| 2017 | 37,938 | 6,233 | 2,321 |
| 2018 | 49,400 | 7,972 | 2,806 |
| 2019 | 66,102 | 11,449 | 4,781 |
| 2020 | 75,355 | 14,016 | 6,126 |
| 2021 | 77,632 | 14,596 | 6,307 |
| 2022 | 72,377 | 14,232 | 5,982 |
| 2023 | 77,037 | 17,722 | 8,232 |

To better visualize the structural shifts in co-affiliation practices over the observation period, Figure S1 illustrates the annual proportions of these cohorts.

## 2. Divergent Characteristics and Trend Analysis of RU and UA Co-affiliations

A longitudinal analysis of the ratios presented in Figure S1 reveals significant divergences in the evolution of scholarly networks between the two countries, particularly following major geopolitical shifts.

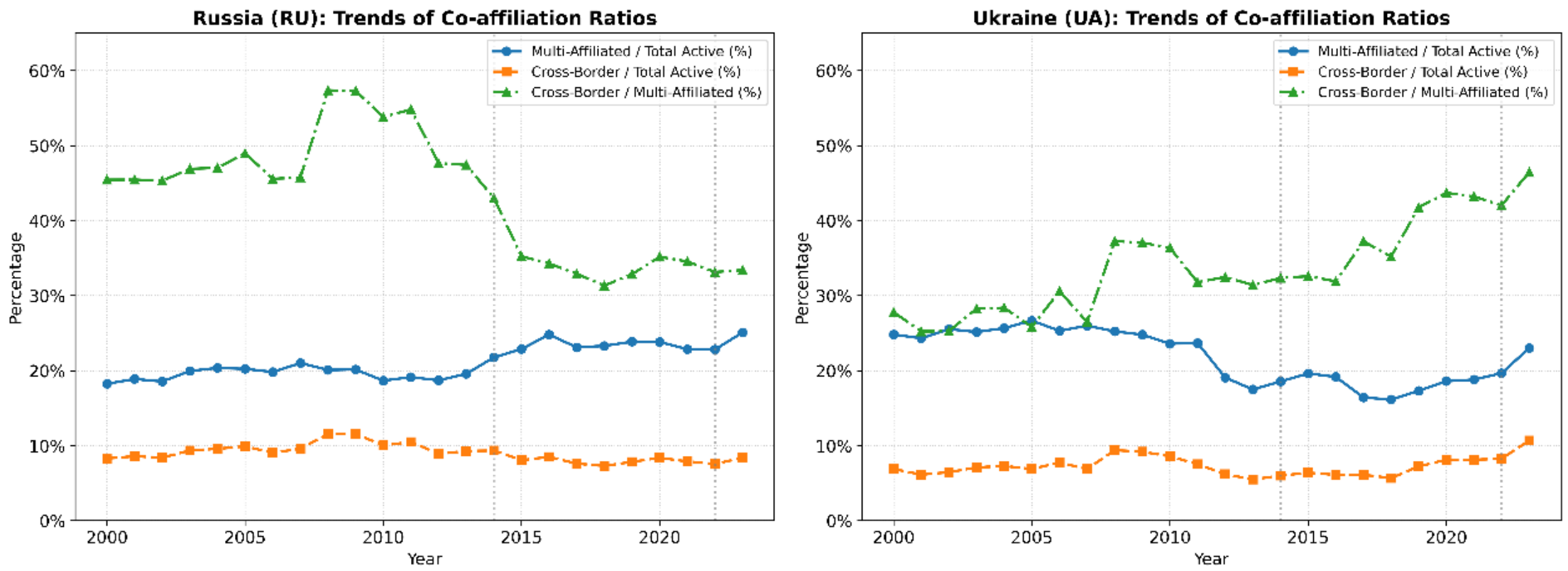

**Figure S1. Comparative Trends of Co-affiliation Ratios for Russian and Ukrainian Scholars (2000–2023).** The blue solid line represents the ratio of multi-affiliated scholars to total active scholars. The orange dashed line denotes the ratio of cross-border affiliated scholars to total scholars. The green dash-dotted line illustrates the proportion of cross-border affiliations within the multi-affiliated cohort. Vertical dashed lines mark the geopolitical events of 2014 and 2022.

### 2.1 Baseline and Evolution of Co-affiliation Proportions

From Convergence to Divergence As indicated by the blue solid line in Figure S1 (Multi-Affiliated Scholars / Total Active), both Russian and Ukrainian scholarly communities exhibited a similar, gradual upward trend in co-affiliation proportions prior to 2014. However, a distinct divergence emerged following the critical juncture of 2022. While the proportion for Ukrainian scholars experienced a sharp positive jump, reaching a peak of approximately 23% in 2023, the Russian proportion plateaued after its 2021 peak, entering a phase of stable fluctuation. This divergence reflects the differing institutional linkage strategies adopted by scholars in response to shifting external environments.

### 2.2 Opposing Trajectories in Cross-Border Structures: Expansion vs. Contraction

The green dash-dotted line (Cross-Border / Multi-Affiliated Scholars) unveils a fundamental contrast in the underlying structure of scholarly networks between the two nations. It is highly noteworthy that the Russian green dash-dotted line exhibits a clear downward trend after 2022. This suggests a contraction in the proportion of international collaborations within the Russian multi-affiliated cohort, with a greater share of co-affiliations occurring between domestic Russian institutions. This inward contraction likely reflects the erosion of transnational academic links due to a combination of international sanctions, academic isolation, and shifts in domestic research priorities toward internal scientific integration.

In stark contrast, the same ratio for Ukraine surged to a historical high of approximately 46.5% after 2022. This outward expansion characterizes a scholarly community where multi-affiliation has become intensely cross-border in nature, indicating that international academic networks have become a primary pillar for sustaining research output amid the disruption of domestic infrastructure.

### 2.3 Dual Anchoring vs. Linkage Dissolution at Critical Junctures

The grey vertical dashed lines in Figure S1 mark the key interventions of 2014 and 2022. In the post-2022 period, the sharp increase in all cross-border ratios for Ukraine (orange dashed and green dash-dotted lines) reflects a strong dual anchoring tendency. Scholars in this cohort appear to be establishing new international institutional ties while simultaneously maintaining their formal affiliations with home institutions.

Conversely, the simultaneous decline in cross-border proportions for Russia suggests a form of linkage dissolution, where the formation of new international ties is insufficient to offset the loss of existing global connections. This comparison provides empirical evidence of the complex behavioral patterns in institutional affiliations under varying degrees of geopolitical pressure.

**3. Methodological Treatment: The Conservative Domestic Anchor Protocol**

The divergent trends observed in Table S5, Table S6, and Figure S1 underscore the necessity of a robust filtering mechanism. Given the prevalence and increasing trajectory of cross-border co-affiliations in both countries—particularly in Ukraine—interpreting any appearance of a foreign institution as a "departure" would lead to a significant overestimation of talent loss due to these complex affiliation patterns.

To mitigate this statistical bias, this study implemented a strict Domestic Anchor Protocol during the data cleaning and classification phases. The guiding principle of this protocol is that the presence of any domestic affiliation indicates that the scholar maintains an observable link with the home country's scientific system.

1. When a scholar's publication record in a given year includes a paper listing both domestic and foreign institutions, the record is interpreted as maintaining an active domestic connection. Consequently, the scholar is not classified as having departed.
2. Furthermore, if a scholar publishes some papers solely with domestic affiliations and others solely with foreign affiliations within the same calendar year, they are still considered to have maintained a domestic presence.
3. A migration event is exclusively triggered and recorded only when a scholar's entire publication portfolio for a given year lists foreign institutions only, with zero observable links to domestic entities.

### 4. Methodological Implications

By transparently disclosing the scale and trend differences of cross-border co-affiliations for both Russia and Ukraine, we acknowledge that institutional belonging in global academic mobility is not a simple binary state but possesses a high degree of complexity.

These rigorous filtering criteria are intentionally designed to preserve the integrity of the longitudinal tracking mechanism. They ensure that the outflow trends captured in this study are driven by a substantive rupture from the domestic scientific infrastructure, rather than statistical noise generated by the expansion of international collaborations or complex part-time appointment statuses. This approach provides a highly conservative and statistically robust lower-bound estimate of the real impact of geopolitical conflict on human capital mobility.

## SUPPLEMENTARY NOTE 4: QUANTITATIVE AUDIT OF AUTHOR DISAMBIGUATION AND PROFILE FRAGMENTATION

### 1. Overview

In large-scale bibliometric research utilizing automated databases such as OpenAlex, the precision of author disambiguation is a critical factor for data integrity. A common challenge in these infrastructures is "profile splitting," where the scholarly output of a single individual is fragmented across multiple unique identifiers (IDs). To ensure the statistical robustness of the longitudinal analysis regarding scholar mobility in Russia and Ukraine, we conducted a systematic quantitative audit to evaluate the extent of profile fragmentation within the target cohort.

### 2. Audit Findings

The audit encompassed all scholarly records associated with Russian and Ukrainian affiliations recorded between 2000 and 2023. We evaluated disambiguation performance by calculating the ratio between unique Author IDs and unique "Display Names" (after standardizing for character encoding, transliteration variants, and naming conventions), as shown in Table S7. This ratio provides a quantifiable measure of potential profile duplication.

**Table S7. Author Disambiguation Metrics for the RU/UA Cohort (2000–2023)**

| Metric | Value |
|---|---|
| Analysis Period | 2000–2023 |
| Target Countries | Ukraine and Russia |
| Total Unique Author IDs | 1,098,389 |
| Total Unique Display Names | 1,047,895 |
| Names Linked to Multiple IDs (Potential Splitting) | 50,494 |
| Estimated Profile Splitting Rate (Upper Bound) | 4.819% |
| Mean IDs per Author Name | 1.048 |

**Note:** The metrics in this table are calculated as follows: (1) Potential Splitting (Excess IDs) = $N_{ID} - N_{Name}$; (2) Profile Splitting Rate (Upper Bound) = $(N_{ID} - N_{Name} / N_{Name}) \times 100\%$; (3) Mean IDs per Author Name = $N_{ID} / N_{Name}$ .

The audit results indicate an estimated profile splitting rate of 4.819%. It is important to emphasize that this figure represents a conservative "upper bound" for algorithmic error. In large populations such as Russia and Ukraine, a substantial portion of these "excess IDs" are attributable to true homonyms—different researchers who happen to share the same name—rather than erroneous system-driven fragmentation.

Consequently, the actual rate of technical profile splitting is likely significantly lower than 4.819%. With a mean of 1.048 IDs per scholar, the data suggests that while micro-level fragmentation may affect individual profiles, the dataset remains highly robust for macro-level analysis. The margin of error is sufficiently low to ensure that aggregate migratory trends and longitudinal patterns are not compromised by disambiguation noise. This quantitative audit confirms that the degree of profile fragmentation within the RU/UA cohort is within an acceptable statistical threshold. The findings support the reliability of the study's conclusions regarding large-scale scholar mobility.

# SUPPLEMENTARY NOTE 5: ROBUSTNESS OF ITS ANALYSIS WITH A ONE YEAR LAG

To assess the robustness of our primary findings regarding the impact of the Russo-Ukrainian conflict on scholar migration, we conducted an Interrupted Time Series analysis with a one-year lag. This adjustment accounts for the potential delay between the conflict event (the annexation of Crimea in 2014) and the subsequent changes in scholars' institutional affiliations as reflected in their publications. Given the inherent lag in the publication process, analyzing the data with a one-year delay provides a more temporally sensitive measure of migration patterns in response to these shocks.

The results of this lagged ITS analysis are presented in Table S8. The table mirrors the structure of the main analysis, presenting the coefficient for the post-shock period (*Coef post shock*), the coefficient for the time trend after the shock (*Coef time after*), and the R-squared value for both Ukraine (UA) and Russia (RU), for the shock year 2014. The corresponding $p$-values for the coefficients are also reported in parentheses. As shown in Table S7, our results support the impact of the conflict on scholar outflow as discussed in the main text. For Ukraine, the 2014 shock exhibits a negative, albeit statistically insignificant ($p$=0.269), immediate impact on scholar outflow, followed by a significant positive trend in the years after (*Coef time after* = 406.720, $p$=0.000). Similarly, for Russia, the 2014 shock presents a negative and statistically insignificant immediate effect on outflow ($p$=0.302), while the subsequent time trend is significantly positive (*Coef time after* = 781.689, $p$=0.000). Overall, the one-year lagged ITS analysis reinforces the conclusions drawn from the main analysis, suggesting that the observed shifts in scholar outflow from both Ukraine and Russia are robust to the temporal lag inherent in publication-based migration data.

**Table S8. ITS analysis of scholar outflow after the conflict with one year lag.**

| Country | Scholar | Coef post shock | Coef time after | R-squared |
|---|---|---|---|---|
| Ukraine | Outflow | -526.238<br>($p$=0.269) | 406.720<br>($p$=0.000) | 0.758 |
| Russia | Outflow | -690.908<br>($p$=0.302) | 781.689<br>($p$=0.000) | 0.877 |

# SUPPLEMENTARY NOTE 6: ANALYSIS OF AUTHORSHIP SCALE AND METHODOLOGICAL THRESHOLDS

**1. Characteristics of Authorship Distribution**

To ensure the empirical rigor of this study, we conducted a systematic review of the authorship scale across the analyzed publication corpus. As illustrated in the provided distribution charts, the number of authors per paper for both Ukrainian and Russian cohorts follows a classic long-tail distribution. The vast majority of research output is produced by small-to-medium-sized collaborative groups, with the "1–5" author category representing the most frequent collaborative mode. This distribution pattern confirms that the dataset primarily reflects standard academic practices common across most scientific disciplines.

**2. Observation of Hyper-authorship**

Our data also captures the phenomenon of hyper-authorship, where a single publication involves an exceptionally high number of contributors. For instance, within the observed window, our dataset identifies a total of 1,300 publications for the Ukrainian cohort and 2,309 publications for the Russian cohort that involve more than 1,000 authors, as shown in Figure S2. These "mega-collaboration" papers are frequently situated in specific research environments, such as high-energy physics consortia or international genomic projects. While these works represent significant collective milestones, their statistical weight under a binary whole-counting approach can be disproportionate. Including them in a country-level network analysis might inadvertently prioritize the institutionalized output of a few massive international consortia over the broader, more sensitive shifts in the general academic community.

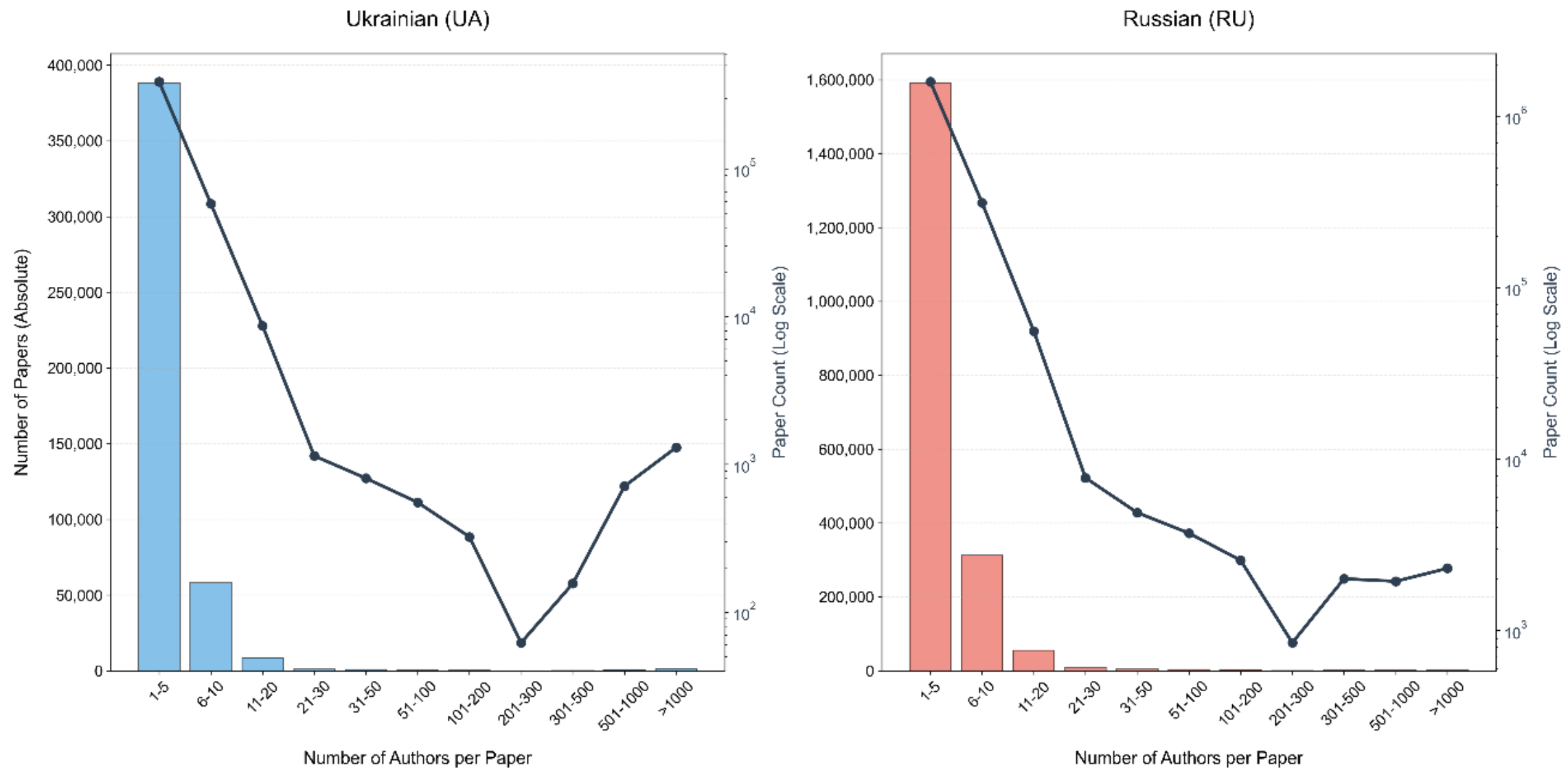

**Figure S2. Observation of Hyper-authorship.**

To maintain a balanced interpretation of international collaboration dynamics, we adopted a filtering threshold of 100 authors. Publications exceeding this limit were excluded from the main analytical framework. This methodological choice was intended to prioritize the observation of "typical" collaborative behaviors—those more likely to be influenced by individual mobility, personal academic networks, and regional geopolitical shifts. By focusing on publications with fewer than 100 authors, the analysis effectively reduces the noise generated by extreme outliers, thereby providing a more representative view of how the majority of scholars in Ukraine and Russia have adapted their collaborative strategies.

The exclusion of hyper-authored papers does not imply a lack of scientific value in those works, but rather serves as a necessary step to stabilize the statistical indicators used in this longitudinal study. This approach ensures that the observed reorientation of research networks, as discussed in Section 4.3, reflects a broad-based structural change within the academic systems. Ultimately, this filtering process enhances the clarity of the findings by ensuring that the reported trends are not driven by a marginal fraction of large-scale experimental projects, but by the wider landscape of scientific collaboration.

# SUPPLEMENTARY NOTE 7: PLACEBO TESTING USING THE DOUBLY ROBUST DIFFERENCE-IN-DIFFERENCES (DRDID) METHOD

To assess the impact of conflict on changes in research topics among Russian and Ukrainian scholars, we employ the Doubly Robust Difference-in-Differences (DRDID) methodology to evaluate the robustness of our data analysis. We construct a placebo dataset and design a detailed estimation strategy. Our analysis is based on a country-topic-publication count panel spanning from 2000 to 2022. To create a credible comparison structure, we randomly assign scholars within each subgroup to treatment and placebo groups. These placebo groups simulate treatment effects in the absence of actual exposure to conflict, enabling us to conduct a placebo test. In other words, we construct hypothetical groups of Ukrainian and Russian scholars presumed unaffected by conflict to compare against the actual treatment groups from both countries.

We begin by matching each scholar's publications to corresponding research topics. For each topic, we aggregate the number of publications per year and calculate the average number of publications in the pre-treatment period (2000–2013) and the post-treatment period (2015–2022). The year 2014, which marks the onset of the crisis, is excluded to avoid contamination from the transition period. The resulting dataset has a typical 2×2 DRDID structure, with two observations—pre- and post-treatment—for each unique combination of country, migration status, treatment assignment, and topic. Each row in the dataset contains the following variables: topic ID (tid), time indicator (time, where 0 indicates pre-treatment and 1 indicates post-treatment), treatment assignment (treat), average number of publications in the period (topic_counts), country (country, coded as UA or RU), and scholar migration status (status, coded as departed or stayed).

For estimation, we apply the DRDID estimator proposed by Callaway and Sant'Anna (2021), implemented using the *drdid* command. Given the repeated cross-section structure of our data, we use the doubly robust inverse probability weighted (DRIPW) estimator, which combines propensity score weighting and outcome regression. We cluster standard errors at the topic level (cluster (topic id)) to account for intra-topic correlation. Separate DRDID estimations are performed within each subsample—Ukrainian and Russian scholars—based on their assigned placebo or treatment status.

As shown in Table S9, results for Ukrainian scholars reveal no statistically significant effects, consistent with expectations under a placebo test. This indicates that the random treatment assignment does not induce systematic differences in topic distributions across groups before and after the conflict, thus confirming that the model does not produce spurious effects in untreated samples. The placebo treatment effects fluctuate around zero, with standard errors within reasonable bounds and *p*-values far from conventional significance thresholds. Table S10 presents results for Russian scholars. The absence of statistically significant treatment effects further supports the conclusion that the distribution of research topics remains stable across groups in the absence of actual conflict exposure. These findings bolster the credibility of our main analysis by affirming the validity of the DRDID model under placebo conditions.

**Table S9. DRDID results for Ukraine scholars.**

| Doubly robust difference-in-differences | | | | | Number of obs = 29,569 | |
|---|---|---|---|---|---|---|
| Outcome model : least squares | | | | | | |
| Treatment model: inverse probability | | | | (Std. err. adjusted for 4,386 clusters in topic ids) | | |
| ATET | Coefficient | Std. err. | z | P>z | [95% conf. interval] | |
| treat (1 vs 0) | 0.835 | 0.572 | 1.460 | 0.144 | -0.029 | 1.957 |

**Table S10. DRDID results for Russian scholars.**

| Doubly robust difference-in-differences | Number of obs = 34,082 |
|---|---|
| Outcome model : least squares | |

| Treatment model: inverse probability | | | | | (Std. err. adjusted for 4,499 clusters in topic ids) | |
|---|---|---|---|---|---|---|
| ATET | Coefficient | Std. err. | z | P>z | [95% conf. interval] | |
| treat (1 vs 0) | -1.119 | 1.422 | -0.790 | 0.431 | -3.905 | 1.667 |

# SUPPLEMENTARY NOTE 8: DESTINATION DISTRIBUTION OF UKRAINIAN AND RUSSIAN SCHOLARS DISPLACED IN 2022

Figure S3 illustrates the primary destinations of scholars emigrating from Ukraine following the full-scale escalation of the Russo-Ukrainian conflict in 2022. In this visualization, the blue node situated within Ukraine represents the origin of the scholars' departure, while the purple arcs delineate their migratory pathways, culminating in the deep blue nodes that signify the receiving countries. The size of these destination nodes indicates the number of scholars received. As depicted, the migration patterns of Ukrainian scholars in 2022 predominantly directed towards neighboring European nations, with Poland, Germany, and the Czech Republic emerging as significant host countries accommodating a substantial number of academics. In comparison to the broader migration trends of both Ukrainian and Russian scholars following the 2014 Crimean conflict as discussed in the main text, the destinations of Ukrainian scholars in 2022 appear to maintain a focus on proximate European states. Notably, Russia continued to be a significant recipient of Ukrainian scholars in 2022, a trend potentially reflecting the enduring influence of geographical, cultural, and linguistic affinities, as well as the return migration of ethnic Russian scholars from Ukraine, factors elaborated upon in the main body of this paper. Furthermore, a proportion of Ukrainian scholars relocated to other European countries and to nations in North America and Asia, including the United States, Canada, China, and Japan.

**Figure S3. Destination distribution of Ukrainian scholars displaced in 2022.**

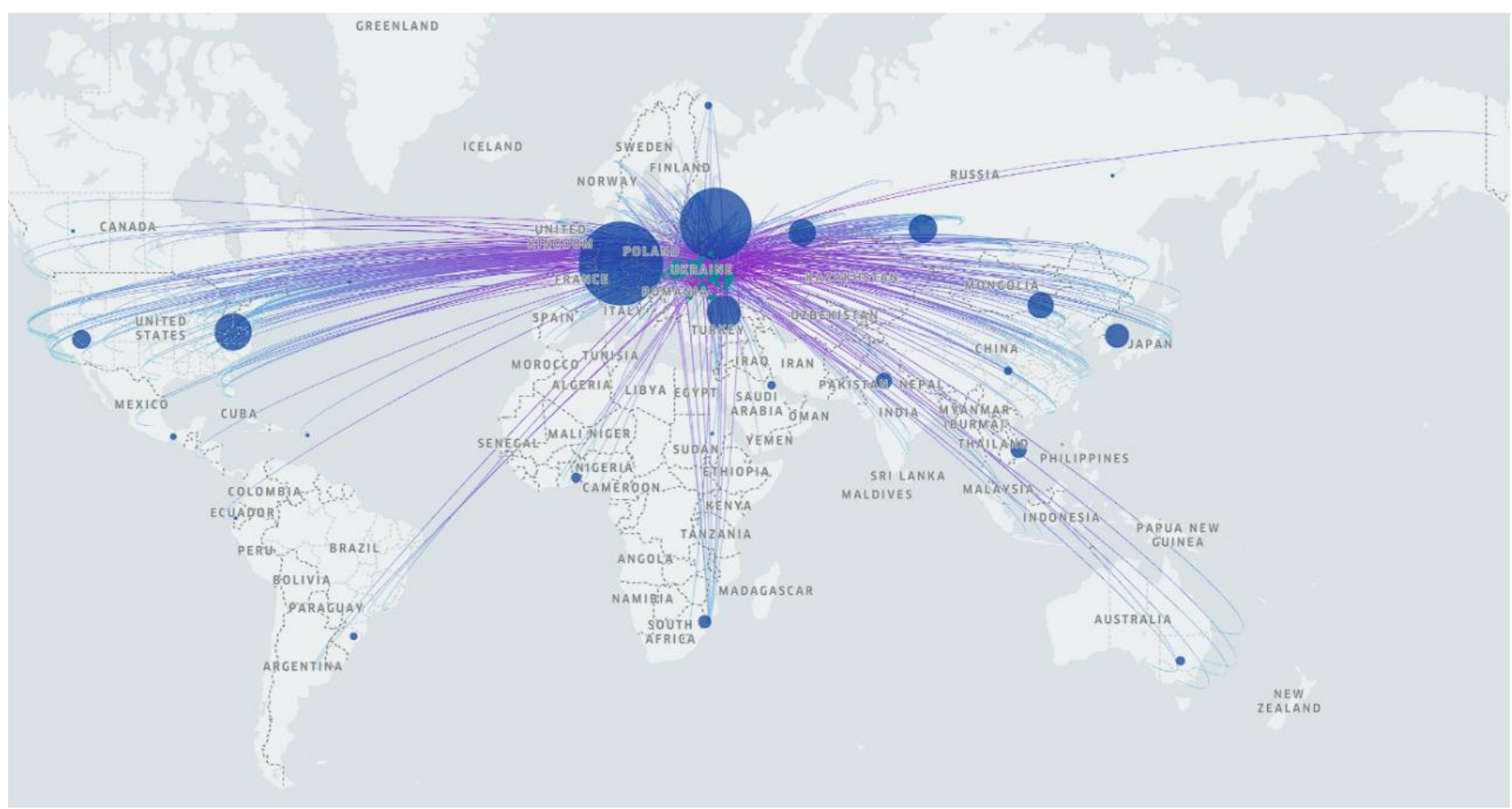

Figure S4 illustrates the primary destinations of scholars emigrating from Russia following the full-scale escalation of the Russo-Ukrainian conflict in 2022. In this visualization, the blue node situated within Russia represents the origin of the scholars' departure, while the blue arcs delineate their migratory pathways, culminating in both orange and light blue nodes that signify the receiving countries. The size of these destination nodes indicates the number of scholars received. As depicted, the migration patterns of Russian scholars in 2022 exhibit a more geographically dispersed distribution compared to their Ukrainian counterparts. While several European countries, including Germany and potentially other nations in Western Europe, received Russian scholars, significant proportions also migrated to destinations across multiple continents. Notably, substantial numbers of Russian scholars relocated to North American countries, particularly the United States and Canada, and to Asian countries, with China and other East Asian nations serving as prominent destinations. Given that the emigrating scholars had at least one scientific publication prior to their departure, it suggests that despite widespread sanctions against the Russian academic community, these measures may be selective in their impact. In other words, Russian scholars with a higher research output may have been conditionally accepted and integrated into research endeavors by numerous countries.

**Figure S4. Destination distribution of Russian scholars displaced in 2022.**

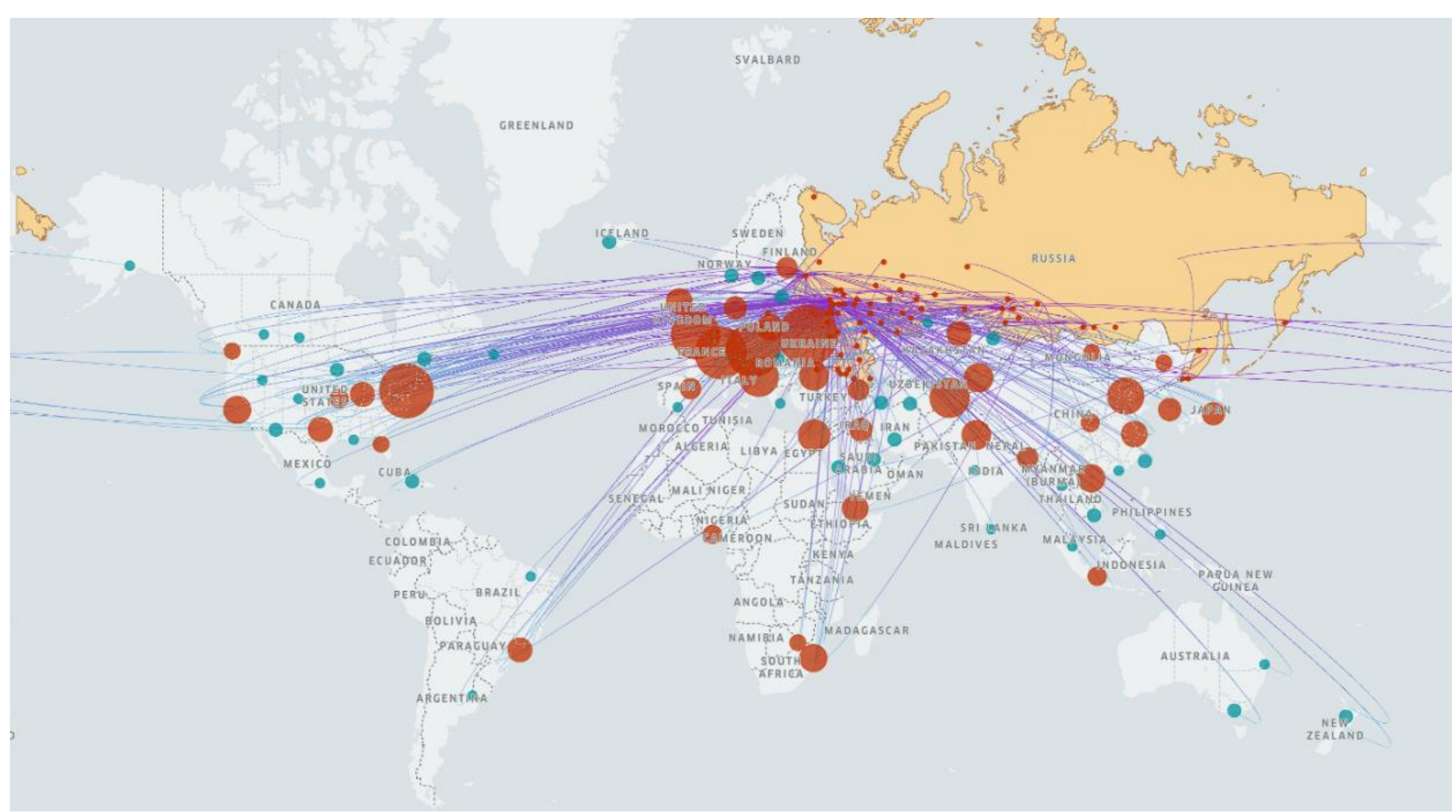

# SUPPLEMENTARY NOTE 9: LONGITUDINAL AND DISCIPLINARY ANALYSIS OF COLLABORATION REORIENTATION ACROSS GEOPOLITICAL PHASES

**1. Introduction**

While Section 4.3 of the main text highlights the macro-level impact of geopolitical conflicts on the international academic collaboration networks of Russian and Ukrainian scholars, aggregate-level metrics may obscure underlying disciplinary heterogeneity. To provide a more granular empirical foundation, this supplementary note utilizes a dataset covering all 26 major scientific disciplines to track scholars' annual international collaboration shares. Direct collaborations between Russia and Ukraine were excluded from this analysis to more clearly highlight the reorientation toward third-party nations. For visual clarity, the trend visualizations are organized alphabetically across two comprehensive panels: Figure S5 displays the first 15 disciplines (from Agricultural and Biological Sciences to Health Professions), and Figure S6 presents the remaining 11 disciplines (from Immunology and Microbiology to Veterinary).

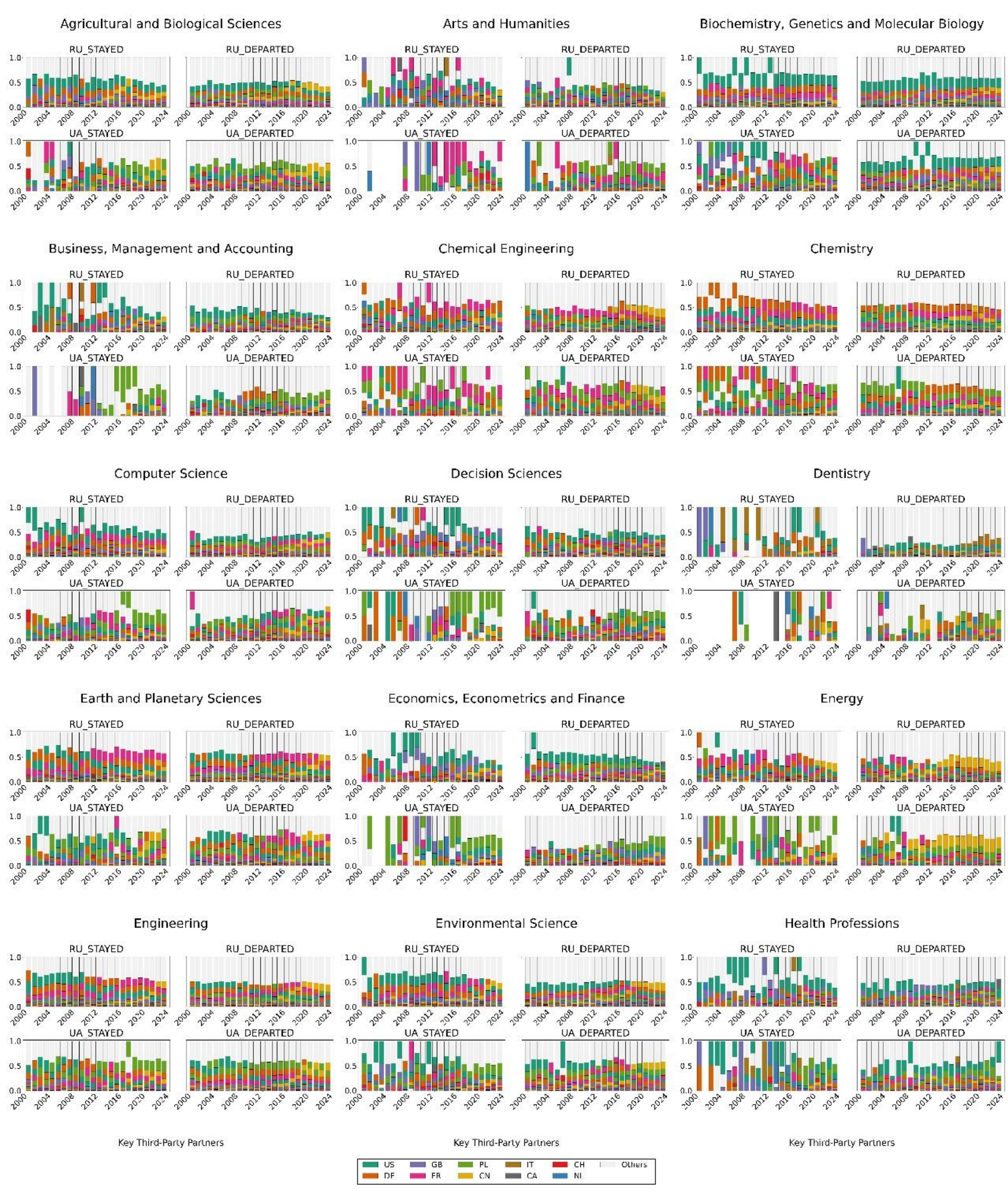

**Figure S5. Annual international collaboration shares for disciplines from Agricultural and Biological Sciences to Health Professions.**

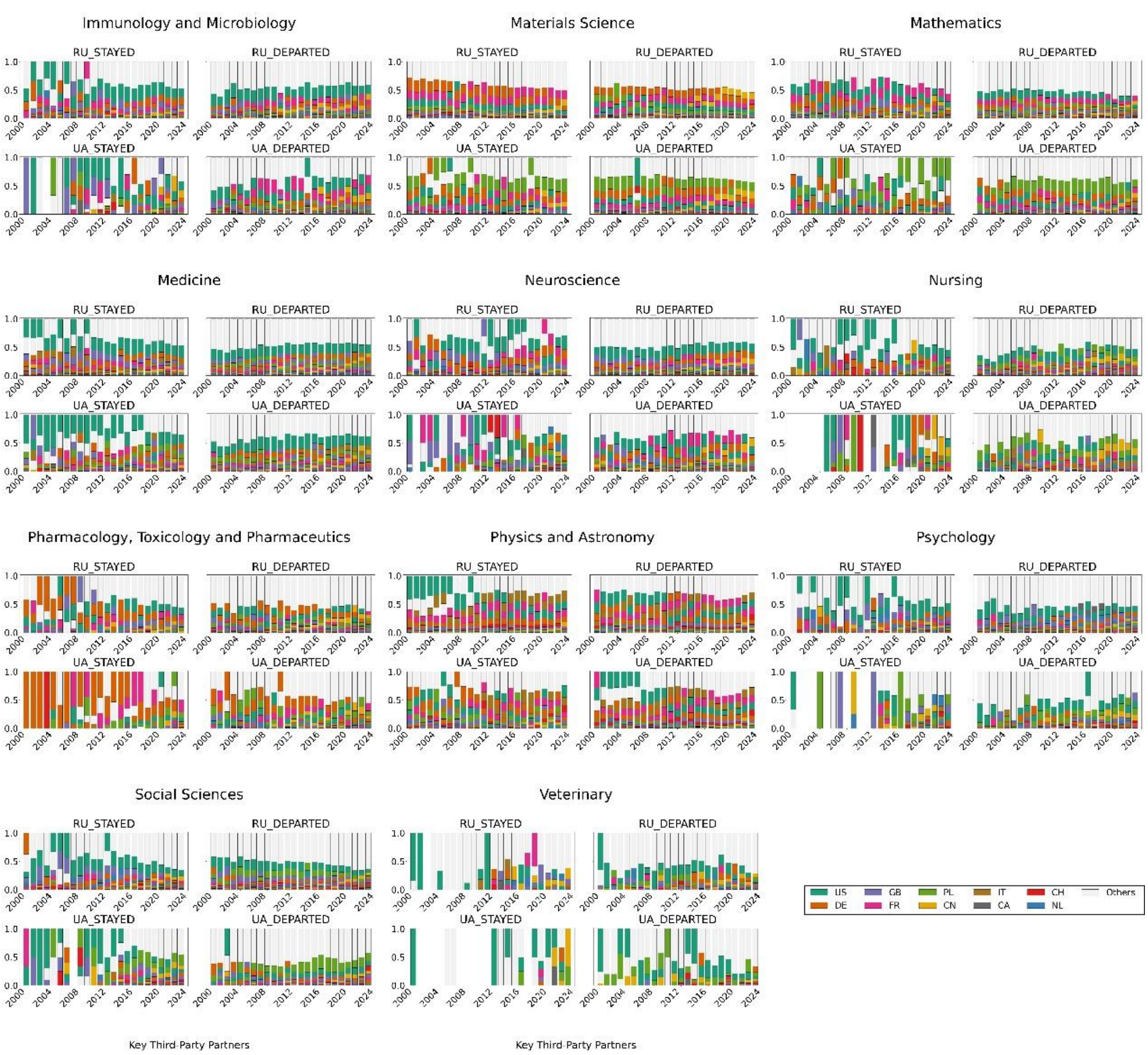

**Figure S6. Annual international collaboration shares for disciplines from Immunology and Microbiology to Veterinary.**

By tracking the share variations of key partner countries, the Western European national matrix, and the "Others" category (representing a broader global network), we systematically illustrate the complex evolution of network reorganization.

## 2. Disciplinary Heterogeneity and Structural Reorientation of Scientific Networks

### 2.1 Russia: The "Eastward Shift" in Applied Hard Sciences and Network Fragmentation in Soft Sciences

The data indicates that geopolitical tensions have had a notable, yet asymmetric, decoupling effect on scholars remaining in Russia (RU_stayed).

In applied and hard science fields, a marked shift in collaborative networks toward non-Western countries (particularly China) is observable. In Materials Science (see Figure 6), the average collaboration share with Germany (DE) gradually decreased from 18.6% (2010–2013) to 16.2% (2014–2021), further declining to 10.5% after 2022. Concurrently, the collaboration share with China (CN) increased from 2.0% before 2014 to 9.2% post-2022. Similar shifts occurred in Engineering (see Figure S5), where the collaboration share with China rose from 1.4% to 12.9%. This reflects a structural adaptation where Russian hard sciences likely sought alternative collaborative partnerships amidst growing restrictions on high-tech equipment and technologies.

However, in fields such as Social Sciences (see Figure S6), where political sensitivities may be higher, these geopolitical pressures translated into pronounced network fragmentation. In this domain, the collaboration share of RU_stayed with the United States (US) decreased from 24.7% (pre-conflict) to 12.0% post-2022. Rather than shifting solely toward China (which increased moderately to 7.5%), these scholars appeared to diversify into a highly dispersed peripheral network—the collaboration share of the "Others" category grew substantially from 37.5% to 64.4%.

In contrast, fundamental "Big Science" disciplines, which rely heavily on cross-national infrastructure, displayed considerable resilience. In Physics and Astronomy (see Figure S6), the collaboration share between RU_stayed and the US remained highly stable across the three phases (17.0% → 14.6% → 14.6%). Furthermore, scholars who departed Russia (RU_departed) did not exhibit a noticeable eastward shift in any discipline, maintaining steady collaboration proportions with countries like the US and Germany. This suggests that the observed decoupling is largely driven by institutional and state-level barriers rather than shifts in individual scholars' academic preferences.

2.2 Ukraine: The Rising Prominence of Western Europe and the Establishment of Proximity

Sanctuaries

For Ukrainian scholars (both stayed and departed), consecutive years of conflict have catalyzed a structural reorientation of their collaborative networks toward European partners.

First, Poland (PL), given its geographical and cultural proximity, established itself as a critical academic hub. For scholars remaining in Ukraine (UA_stayed), deteriorating domestic research conditions increased their reliance on knowledge production spillovers from neighboring countries. In Computer Science (see Figure S5), Poland's collaboration share doubled from 12.4% (pre-2014) to 24.7% (2014-2021), reaching 27.3% after 2022. For departed scholars (UA_departed), Poland also provided a stable and significant share (14% - 16%) in Materials Science (Figure S6) and Engineering (Figure S5).

Second, the collaborative positions of Western European countries (e.g., Germany, the UK, France, Italy) and the US experienced a systemic elevation. The data suggests that the Euro-Atlantic academic community not only hosted displaced scholars but also provided targeted research support to those remaining in Ukraine. In Physics and Astronomy (see Figure S6), despite the impact of the conflict, the collaboration share of UA_stayed with the US grew from 8.6% (pre-2014) to 11.9% post-2022. In Medicine, the collaborative matrix representing Western European countries like Germany (DE) and the UK (GB) notably expanded.

Finally, the large baseline of the "Others" category reveals the diversification of Ukraine's support networks. In the data for departed Ukrainian scholars (UA_departed), the "Others" category in Agricultural and Biological Sciences (see Figure S5) reached 47.8% after 2022. Unlike the fragmentation observed in Russian social sciences, the high share of "Others" here likely reflects Ukrainian scholars integrating into a vast array of pan-European academic institutions and non-traditional global nodes. Concurrently, Ukraine's collaboration share with China across the disciplines remained marginally low (generally below 6%), supporting the observation that the Ukrainian research network has become broadly integrated into a diversified, Western-oriented academic alliance.

## 3. Spatial Reconfiguration and Absolute Scale of Regional Realignment

The net change in publication counts between 2021 and 2023, as illustrated in Figure S7, offers a more direct observation of the spatial reorganization of research networks. This shift in absolute values reveals a profound divergence in the regional collaborative strategies adopted by different scholar cohorts in response to geopolitical shocks. Unlike percentage-based metrics, these absolute counts highlight the physical scale of the transition from historical dependencies to new regional anchors.

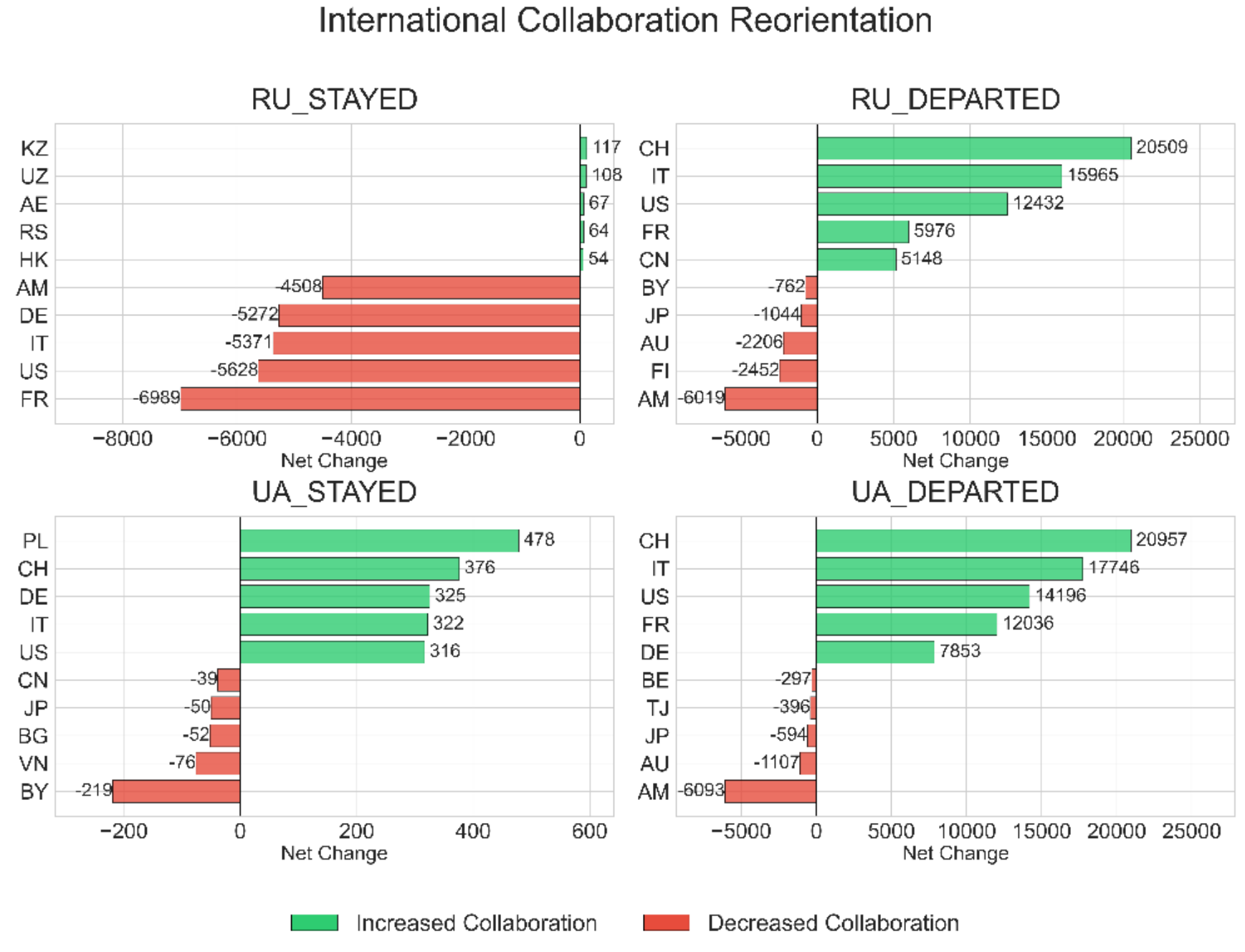

**Figure S7. Net change in raw international collaboration counts across scholar cohorts by partner country.**

For scholars remaining in Russia (RU_stayed), the absolute decline in collaboration with traditional Western scientific hubs is a defining characteristic of this period. Significant decreases in collaborative volume are recorded with the United States (US), Germany (DE), United Kingdom (GB), France (FR), and Italy (IT). This reduction indicates a contraction of the institutional ties that previously linked the Russian domestic research system to the Western core. In contrast, a compensatory "Eurasian pivot" is clearly visible through the growth in partnerships within the Commonwealth of Independent States and East Asia. Beyond the substantial growth with China (CN), there is a notable increase in collaborative output with Central Asian nations, specifically Kazakhstan (KZ) and Uzbekistan (UZ). This suggests that Russian institutions are increasingly leveraging regional, linguistic, and historical ties to mitigate international isolation. Furthermore, the absolute growth in ties with India (IN) and Turkey (TR) reflects a strategic diversification toward emerging scientific powers in the Global South, attempting to establish a multi-polar collaborative model outside traditional Western channels.

The collaborative profile of migrated Russian scholars (RU_departed) presents a starkly different spatial footprint, closely aligning with the known corridors of intellectual emigration since 2022. This cohort shows significant absolute growth in collaborations with specific intermediary hubs, most notably Armenia (AM), Georgia (GE), Serbia (RS), Israel (IL), and the United Arab Emirates (AE). These countries have effectively become central nodes for relocated academic capital, maintaining logistical and professional accessibility for displaced researchers. At the same time, this group has maintained and even expanded its collaborative links with Western nations such as Germany (DE), the United States (US), and the United Kingdom (GB). This divergence from the RU_stayed cohort suggests that once Russian scholars relocate outside their domestic institutional framework, they are successfully assimilated into Western-centric or intermediary research networks.

For scholars remaining in Ukraine (UA_stayed), the data reveals a rapid and definitive structural break from the legacy post-Soviet scientific ecosystem. The most substantial absolute decreases in collaborative volume are observed with Russia (RU) and Belarus (BY). Given its role as a key strategic ally of Russia, the severing of ties with Belarus represents a comprehensive withdrawal from the Moscow-led regional scientific bloc. This decoupling is replaced by a pivot toward a Central-Eastern European corridor. Absolute gains in collaborative output are concentrated in Poland (PL), Czechia (CZ), and Germany (DE), alongside targeted growth with the United States (US). Additionally, increased ties with the Baltic states, including Estonia (EE) and Latvia (LV), as well as Slovakia (SK), reinforce the establishment of a regional academic sanctuary. This realignment provides empirical evidence that Ukraine's domestic research system has transitioned from historical regional dependencies to a model deeply integrated with its Western and European neighbors.

Migrated Ukrainian scholars (UA_departed) exhibit an even broader integration into the European Research Area and the transatlantic scientific community. This cohort shows a significant absolute increase in publications involving a wide array of European partners, including Poland (PL), Germany (DE), the United Kingdom (GB), Spain (ES), Czechia (CZ), France (FR), and Italy (IT). This broad-based growth demonstrates that displaced Ukrainian researchers are not merely seeking temporary refuge but are being systematically integrated into the core of the Western scientific establishment. The concurrent rise in collaborative volume with the United States (US) further underscores the robust capacity of the international academic community to absorb and support Ukrainian talent during the conflict. These findings confirm a total and diversified structural realignment of the Ukrainian scientific future away from its former regional ties.

## 4. Conclusion

Longitudinal data analysis covering 26 core disciplines and two mobility statuses suggests that geopolitical factors have induced a multi-dimensional structural reorganization of research networks rather than a uniform global scientific fracture. At the disciplinary level, the findings indicate that Russian applied hard sciences exhibit a tendency toward an eastward shift, while its social sciences have experienced notable network fragmentation. Conversely, the Ukrainian research community shows a trend toward deeper integration into the Euro-Atlantic academic network, supported by collaborative assistance from Western Europe and the United States.

Beyond these structural shifts, the analysis of absolute publication volumes reveals a reconfiguration of scientific alliances in terms of spatial distribution. The data suggests that the Russian domestic research system appears to be establishing new regional collaborative anchors in Central and East Asia—most notably with Kazakhstan, Uzbekistan, and China—while diversifying toward emerging scientific powers in the Global South. Simultaneously, the Ukrainian scientific community exhibits characteristics of decoupling from the legacy post-Soviet ecosystem, specifically reducing collaborative ties with Russia and Belarus, and realigning toward a Central-Eastern European regional network comprising Poland and the Baltic states.

Furthermore, the divergence between stayed and departed cohorts suggests that while institutional factors drive the adjustment of collaborative relationships at the state level, individual mobility facilitates the continued integration of human capital into the global mainstream. These multi-disciplinary empirical findings reflect a strategic adjustment of scientific networks. This reorganization suggests that research networks are evolving from historical regional dependencies toward a more diversified global research landscape that is increasingly integrated with Western scientific systems.

## SUPPLEMENTARY NOTE 10: QUANTITATIVE CONTEXT OF HOST COUNTRY RESEARCH CAPACITY AND SCALE DIFFERENCES

To provide a more comprehensive context for interpreting authorship role transitions among migrated scholars, we analyzed the academic scale and research capacity of the major destination countries within our dataset. As illustrated in Figure S8, there is a significant longitudinal variance between 2000 and 2023 in the annual number of unique active authors and publication volumes across these nations. Both metrics are presented on a log scale to visualize the multi-order magnitude differences in research system scale among host nations without implying a hierarchy of research quality.

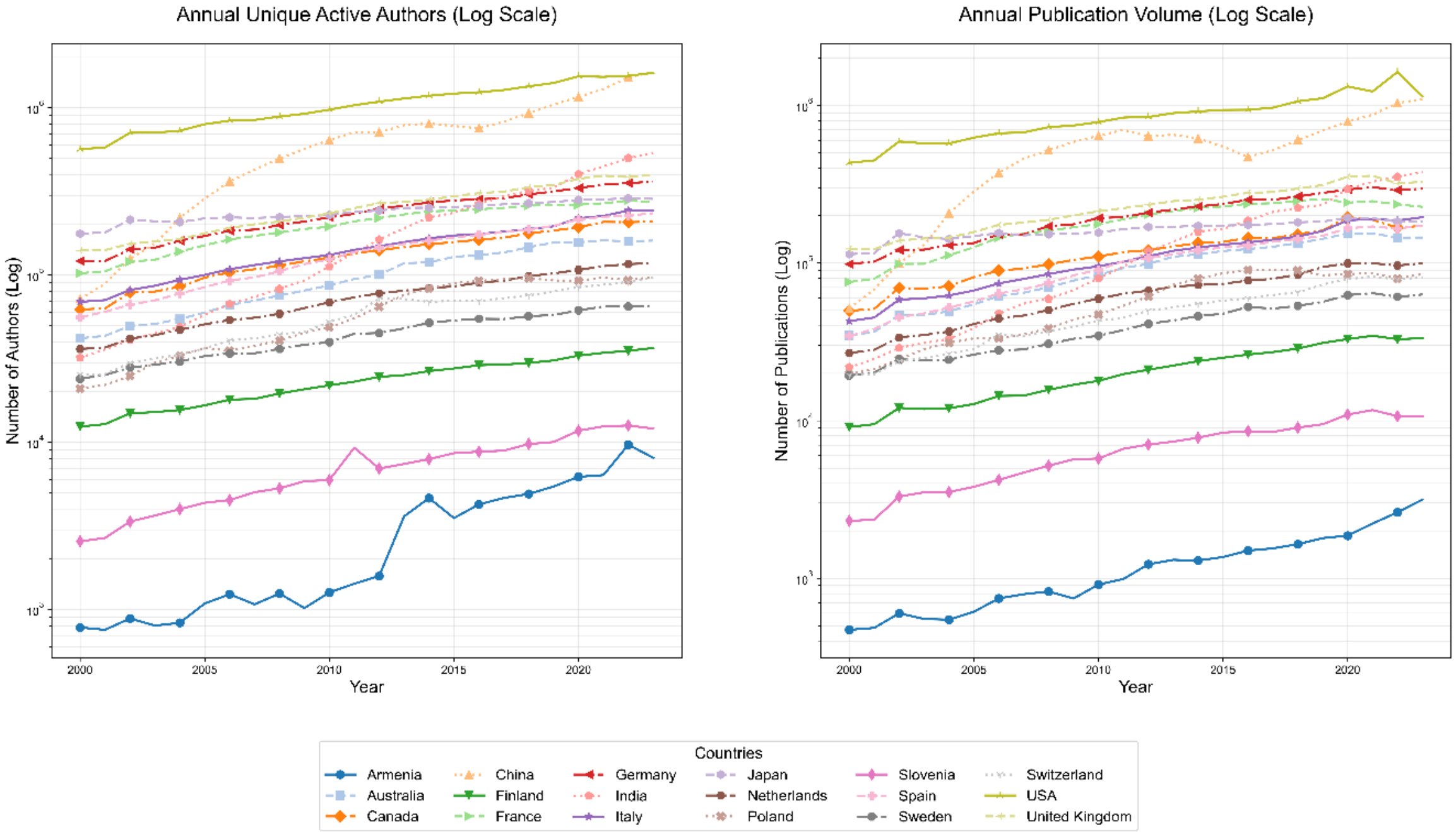

**Figure S8. Comparative analysis of host country research capacity.** The left panel displays the annual count of unique active authors, and the right panel displays the annual publication volume for 18 destination countries. Both metrics are plotted on a log scale to illustrate the variance in academic infrastructure and research output scale.

The scale of these research systems can be quantified by comparing the capacity of various host countries for the representative year of 2023. In that year, the United States recorded 1,144,092 annual publications and supported a population of 1,617,943 active authors. Similarly, China reached a volume of 1,096,973 publications with 1,637,270 authors, and Germany recorded 296,453 publications and 361,263 authors. These figures suggest an environment characterized by a high density of active researchers and established institutional structures.

In contrast, during the same year (2023), Armenia recorded 3,156 publications and 8,065 authors, while Slovenia had 10,645 publications and 12,061 authors. These underlying numbers indicate that the research systems of some host countries are several orders of magnitude different in scale compared to others. For instance, the active author population in the United

States was approximately 200 times larger than that of Armenia in 2023.

This variance in research system scale may offer a relevant context for understanding scholarly integration. In systems with a relatively smaller number of active researchers and lower overall publication volumes, migrated scholars might encounter a different set of institutional conditions. Such environments could potentially facilitate the identification of a scholar's expertise and might be associated with different pathways toward research leadership roles.

Conversely, the patterns observed in larger scientific systems, where a decline in leadership roles was more consistent among certain migrated cohorts, may be interpreted in light of the high density of researchers and the complex institutional hierarchies inherent in large-scale infrastructures. By contextualizing authorship positions within these baseline capacity metrics, we aim to provide a more balanced assessment of the professional adjustments associated with migration, acknowledging that the structural characteristics of the host country represent one of several potential factors influencing academic role transitions.

## SUPPLEMENTARY NOTE 11: MULTIDISCIPLINARY ANALYSIS OF AUTHORSHIP ROLE TRANSITIONS

To ensure the empirical validity of the findings regarding authorship role transitions, this multidisciplinary analysis was conducted using a systematic selection of the ten most represented disciplinary categories within the dataset. This selection strategy was intentionally designed to mitigate disciplinary composition bias, a potential distortion where aggregate shifts in authorship roles might be misattributed to the impact of migration, when they are, in fact, merely reflections of the varying signature norms across different research communities. By isolating the longitudinal trends within these diverse fields—spanning from high-collaboration "Big Science" environments to more individual-centric disciplines—we aimed to avoid methodological noise and to verify whether the observed patterns of academic role adjustments persist as a consistent phenomenon, regardless of field-specific authorship conventions.

The baseline of academic leadership in home countries is clearly established in the panels for stayed scholars within Figure S9 (Russian cohorts) and Figure S10 (Ukrainian cohorts). Across all ten analyzed disciplines, scholars who remained in their home academic systems maintained a remarkably stable and dominant role distribution throughout the 2015–2023 period. In these cohorts, the proportion of first authorship (indicated by the blue line) consistently ranks highest, significantly exceeding both middle (red line) and last authorship (green line) shares. This "first-author dominant" pattern remains robust even following the escalation of tensions in 2022. In experimental sciences such as Chemistry, Engineering, and Materials Science, first-authorship shares for stayed scholars frequently remained above 0.9, reflecting a high degree of research autonomy and leadership within their domestic institutional frameworks.

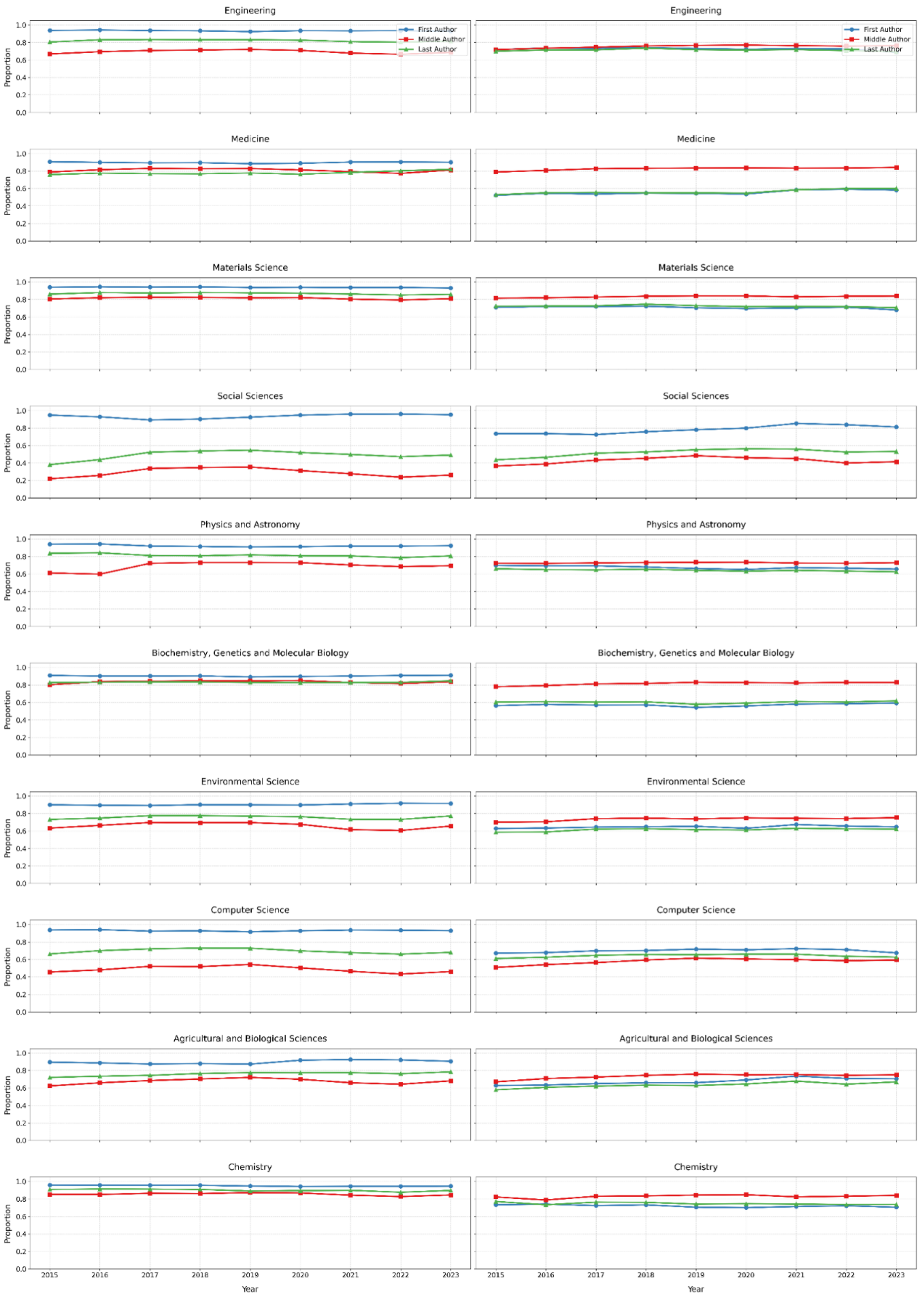

**Figure S9. Annual proportions of authorship roles for Russian stayed and departed scholars across ten major disciplines.**

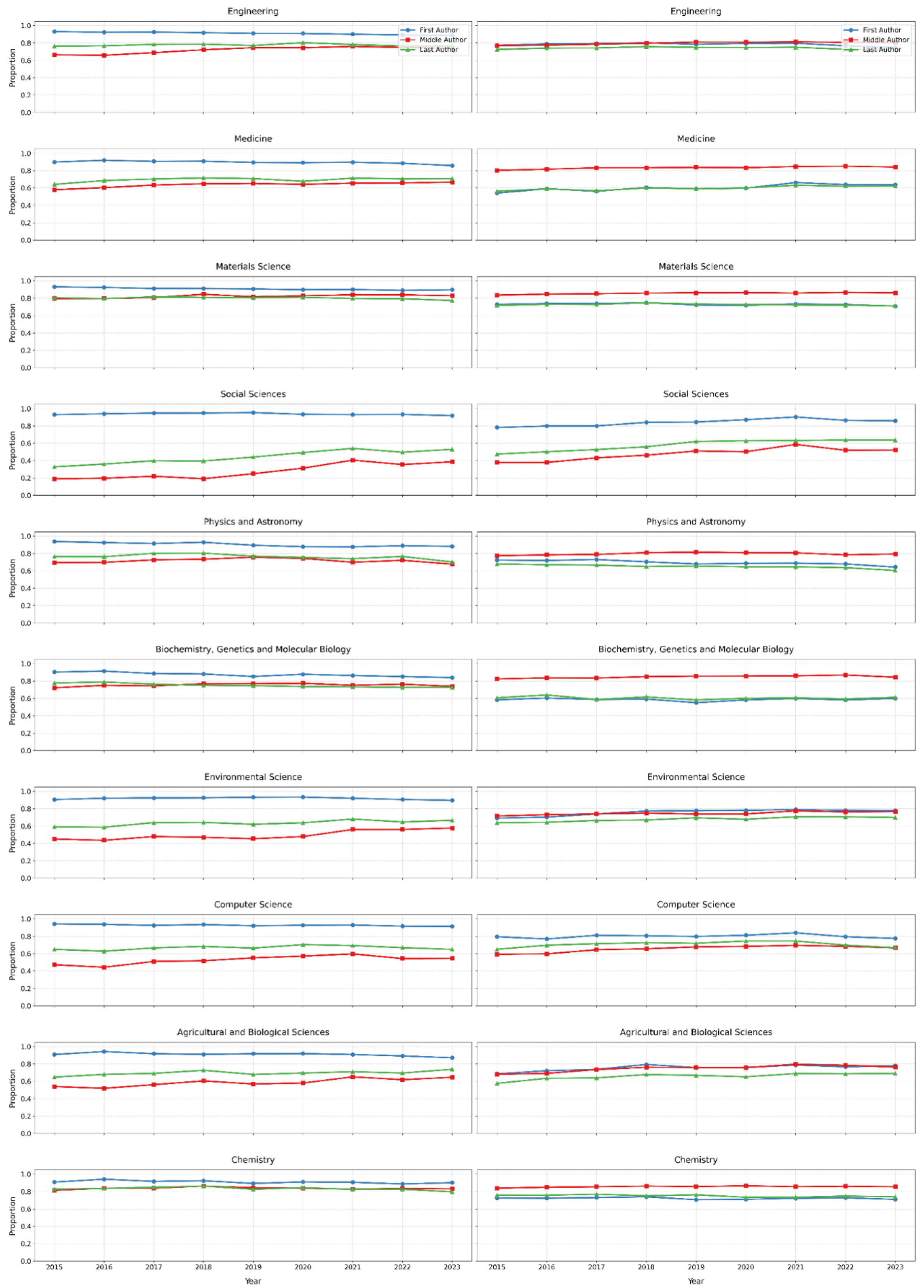

**Figure S10. Annual proportions of authorship roles for Ukrainian stayed and departed scholars across ten major disciplines.**

In contrast, the analysis of departed cohorts in Figure S9 and Figure S10 reveals a structural reorientation of authorship roles following cross-border migration. The most prominent feature of this transition is the "role inversion" phenomenon, where the share of middle authorship rises to surpass or equal that of first or last authorship. Quantitatively, the number of disciplines exhibiting this inversion increased noticeably after the onset of the conflict. For the Russian departed cohort (Figure S9), this shift is observed in the vast majority of the ten analyzed fields—including Engineering, Medicine, Biochemistry, and Materials Science—where the red line representing middle authorship ascended to the dominant position by 2022. Similarly, for the Ukrainian departed cohort (Figure S10), even in fields such as Social Sciences, which traditionally emphasize individual contributions, the middle-authorship proportion showed a visible increase compared to pre-conflict baseline levels. This trend suggests that upon entering new host academic systems, migrated scholars frequently adopt a strategy of joining established research teams as collaborators to ensure research continuity during their initial integration phase.

The longitudinal dynamics following the 2022 escalation further highlight the impact of the external environment on these role transitions. The number of disciplines where middle authorship became the primary role reached its peak in the 2022–2023 period for both departed groups. For the Russian departed cohort in Figure S8, the gap between middle authorship and leadership roles (first and last) widened in disciplines such as Biochemistry and Environmental Science during these years. Conversely, the Ukrainian departed cohort in Figure S10 showed a more varied adaptive trajectory. While middle-authorship shares remained elevated compared to home-country baselines, several disciplines, such as Computer Science, exhibited signs of stabilization or slight recovery in first-author shares. This asymmetry in role evolution likely reflects the differing availability of institutional support mechanisms and the varying degrees of geopolitical pressure faced by each cohort in their respective host countries.

By covering ten diverse disciplines within these two consolidated figures, this analysis effectively addresses potential concerns regarding disciplinary authorship norms. While certain fields, such as Physics and Astronomy, naturally exhibit closer proximity between the three authorship lines due to long-standing hyper-authorship traditions, the relative shift toward middle-authorship dominance remains observable for migrated scholars in both Figure S9 and Figure S10. In disciplines with clearer role boundaries, such as Medicine or Agricultural and Biological Sciences, the displacement from "leader" to "collaborator" is even more pronounced. The cross-disciplinary consistency of these patterns confirms that the observed shifts in authorship roles are primarily driven by the professional adjustments and network reconfigurations associated with migration and geopolitical instability, rather than by shifts in the disciplinary composition of the research output.

In conclusion, the evidence presented in Figure S9 and Figure S10 suggests that while scholars remaining in their home countries maintain a high level of academic leadership, those who migrate undergo a systematic transition from independent leadership toward deeper collaborative integration. The observed inversion of authorship roles—where middle authorship frequently replaces first authorship as the most common position—provides a micro-level indicator of how global academic power and scholarly identity are being redistributed under the influence of regional and global tensions.